\documentclass[12pt,preprint]{aastex}

\def\chandra{{\it Chandra\/}}
\def\conx{{\it Constellation-X\/}}
\def\genx{{\it Generation-X\/}}

\def\hst{{\it {\it HST}\/}}

\def\rosat{{\it ROSAT\/}}

\def\xeus{{\it XEUS\/}}

\def\xmm{{\it XMM-Newton\/}}

\def\ltsima{$\; \buildrel < \over \sim \;$}
\def\simlt{\lower.5ex\hbox{\ltsima}}
\def\gtsima{$\; \buildrel > \over \sim \;$}
\def\simgt{\lower.5ex\hbox{\gtsima}}
\def\kms{\ifmmode{~{\rm km~s^{-1}}}\else{~km s$^{-1}$}\fi}
\def\lsim{\lower0.3em\hbox{$\,\buildrel <\over\sim\,$}}
\def\gsim{\lower0.3em\hbox{$\,\buildrel >\over\sim\,$}}
\def\hst{{\it HST}}
\def\msol{$M_\odot$}
\def\h2{H$_2$}
\def\flux{erg~cm$^{-2}$~s$^{-1}$}
\def\xlum{erg~s$^{-1}$}

\def\arcsec{\mbox{$^{\prime\prime}$}}

\def\Lx{$L_{\mbox{\scriptsize{X}}}$}

\shortauthors{LEHMER ET AL.}
\shorttitle{Off-Nuclear Sources in the Chandra Deep Fields}

\begin{document}


%
\title{The Properties and Redshift Evolution of Intermediate-Luminosity Off-Nuclear X-ray Sources in the \chandra\ Deep Fields}
%

\author{
B.~D.~Lehmer,\altaffilmark{1}
W.~N.~Brandt,\altaffilmark{1}
A.~E.~Hornschemeier,\altaffilmark{2}
D.~M.~Alexander,\altaffilmark{3}
F.~E.~Bauer,\altaffilmark{4}
A.~M.~Koekemoer,\altaffilmark{5}
D.~P.~Schneider,\altaffilmark{1}
\& A.~T.~Steffen\altaffilmark{1}
}
\altaffiltext{1}{Department of Astronomy \& Astrophysics, 525 Davey Lab,
The Pennsylvania State University, University Park, PA 16802, USA}
\altaffiltext{2}{Laboratory for X-ray Astrophysics, NASA Goddard Space Flight Center, Code 662, Greenbelt, MD 20771, USA}
\altaffiltext{3}{Institute of Astronomy, Madingley Road, Cambridge, CB3 0HA, United Kingdom}
\altaffiltext{4}{Columbia Astrophysics Laboratory, Columbia University, Pupin Labortories, 550 W. 120th St., Rm 1418, New York, NY 10027, USA}
\altaffiltext{5}{Space Telescope Science Institute, 3700 San Martin Drive, Baltimore, MD 21218, USA}
%
\begin{abstract}
%

We analyze a population of intermediate-redshift ($z \approx 0.05-0.3$),
off-nuclear \hbox{X-ray} sources located within optically-bright galaxies in
the Great Observatories Origins Deep Survey (GOODS) and Galaxy Evolution from
Morphology and SEDs (GEMS) fields.  A total of 24 off-nuclear source candidates
are classified using deep \chandra\ exposures from the \chandra\ Deep
Field-North, \chandra\ Deep Field-South, and Extended \chandra\ Deep
Field-South; 15 of these are newly identified.  These sources have average
\hbox{X-ray} spectral shapes and optical environments similar to those observed
for off-nuclear intermediate-luminosity (\Lx~$\ge 10^{39}$ \xlum\ in the
\hbox{0.5--2.0~keV} band) \hbox{X-ray} objects (IXOs; sometimes referred to as
ultraluminous \hbox{X-ray} sources [ULXs]) in the local universe.  This sample
improves the available source statistics for intermediate-redshift, off-nuclear
sources with \Lx~$\simgt 10^{39.5}$ \xlum, and it places significant new
constraints on the redshift evolution of the off-nuclear source frequency in
field galaxies.  The fraction of intermediate-redshift field
galaxies containing an off-nuclear source with \Lx~$\simgt$10$^{39}$ \xlum\ is
suggestively elevated ($\approx$80\% confidence level) with respect to that
observed for IXOs in the local universe; we calculate this elevation to be a
factor of $\approx$1.9$^{+1.4}_{-1.3}$.  A rise in this fraction is plausibly
expected as a consequence of the observed increase in global star-formation
density with redshift, and our results are consistent with the expected
magnitude of the rise in this fraction.

%
\end{abstract}
%


%
\section{Introduction}
%

Intermediate-luminosity \hbox{X-ray} objects (IXOs) in the local universe are
off-nuclear \hbox{X-ray} sources having
\hbox{0.5--8.0~keV} luminosities exceeding $\sim$10$^{39}$ \xlum\ (e.g.,
Colbert \& Ptak 2002).  \hbox{X-ray} sources making up the high-luminosity end
(i.e., \hbox{$\simgt$2 $\times$ 10$^{39}$ \xlum}) of the IXO population are commonly
referred to as ultraluminous \hbox{X-ray} sources (ULXs); they are referred to
as ``ultraluminous'' because their inferred isotropic luminosities exceed that
expected for an $\approx$10~\msol\ black hole accreting at the Eddington limit.
Investigations of this enigmatic population suggest that many IXOs may be
intermediate-mass black holes (IMBHs) with masses exceeding $\approx$10~\msol\
(e.g., Colbert \& Miller 2004; Miller et~al. 2004) and/or normal high mass
\hbox{X-ray} binaries (HMXBs) anisotropically beaming \hbox{X-rays} into our
line-of-sight (e.g., King et al.  2001).  Recent observations indicate IXOs may
be a natural high-luminosity extension of the HMXB \hbox{X-ray} luminosity
function, suggesting that these sources trace relatively young stellar
populations (e.g., Gilfanov et~al.  2004).  Indeed, analyses of the physical
and environmental properties of IXOs have shown that the number and cumulative
\hbox{X-ray} luminosities of IXOs are correlated with star-formation rate
(e.g., Kilgard et al. 2002; Swartz et al.  2004).  As a consequence of this,
IXOs are more commonly observed in late-type spiral and irregular galaxies than in
early-type ellipticals.  Furthermore, investigations of \hbox{X-ray} populations in
early-type galaxies find that luminous (\Lx~$\simgt 2 \times 10^{39}$~\xlum)
IXOs (i.e., ULXs) are generally absent (e.g., Irwin et al.  2003); however,
exceptions have been noted (e.g., Loewenstein et al. 2005).

Recently, the fraction of galaxies in the local universe containing IXOs (as a
function of \hbox{X-ray} luminosity) has been constrained statistically using
\rosat\ observations of a sample of 766 relatively nearby ($D < 66.7$~Mpc)
galaxies (Ptak \& Colbert 2004; hereafter PC04) from the {\it Third Reference
Catalog of Bright Galaxies} (RC3; de Vaucouleurs et al. 1991).  PC04 report
that, after correcting for expected background sources, $\approx$12\% and
$\approx$1\% of all local RC3 spiral and irregular galaxies have one or more IXOs with
\hbox{2--10~keV} luminosities \hbox{\Lx\ $> 10^{39}$ erg s$^{-1}$} and
\hbox{\Lx\ $> 10^{40}$ erg s$^{-1}$}, respectively.  Here, \hbox{2--10~keV}
luminosities were estimated from the \hbox{0.2--2.4~keV} flux with
an assumed IXO power-law photon index of $\Gamma=1.7$.

At increasing redshifts, it is plausibly expected that the fraction of galaxies
hosting IXOs will increase as a result of the observed rise in global
star-formation density with redshift (e.g., Madau et al. 1998; Ghosh \& White 2001).  Evidence for
the global evolution of star-formation activity has also been observed as an
increase in the average \hbox{X-ray} luminosity from normal galaxies out to $z
\approx 1$ (e.g., Hornschemeier et al. 2002); IXOs likely play an important
role in this evolution.  Thus far, low \hbox{X-ray} flux levels and limited
angular resolution have restricted the study of relatively distant ($z \simgt
0.05$; $D \simgt 200$~Mpc) IXOs.  However, deep multiwavelength extragalactic
surveys that combine the optical imaging capabilities of the {\it Hubble Space
Telescope} ({\it HST\/}) and the sub-arcsecond \hbox{X-ray} imaging of the
\chandra\ \hbox{X-ray} Observatory (\chandra) have made the detection and
classification of intermediate-redshift ($z \approx 0.05-0.3$; lookback times
of $\approx$ 0.7--3.4~Gyr) off-nuclear sources possible.  Hornschemeier et~al.
(2004; hereafter H04) isolated and characterized ten off-nuclear source
candidates in optically-bright field galaxies with redshifts in the range $z =
0.04-0.23$ ($z_{\rm median} \approx 0.1$).  This investigation utilized \hst\
observations with Advanced Camera for Surveys (ACS) filters $B_{435}$,
$V_{606}$, $i_{775}$, and $z_{850}$ from the Great Observatories Origins Deep
Survey (GOODS; Giavalisco et~al. 2004) and deep \chandra\ observations
coincident with these fields through the $\approx$2~Ms \chandra\ Deep
Field-North (\hbox{CDF-N}; Alexander et~al. 2003) and $\approx$1~Ms \chandra\
Deep Field-South (\hbox{CDF-S}; Giacconi et~al.  2002) surveys.  H04 found that
the fraction of field galaxies with detectable off-nuclear \hbox{X-ray} sources
($L_{\rm 0.5-2.0~keV} \simgt 10^{38.9}$~\xlum) at $z \approx 0.1$ is
$36_{-15}^{+24}\%$, suggestively larger than that observed for galaxies in the
local universe.  Moreover, due to an angular-resolution bias, this ``observed''
fraction was only considered to be a lower limit to the true fraction, which
would include off-nuclear sources with offsets smaller than the \chandra\
positional error circles.  Unfortunately, a study of the dependence of this
fraction upon off-nuclear source \hbox{X-ray} luminosity fraction was not
possible due to limited source statistics, most notably for sources with
\hbox{0.5--2.0~keV} luminosities \Lx~$\simgt 10^{39.5}$~\xlum. 

In this investigation, we estimate the true fraction of intermediate-redshift
field galaxies hosting off-nuclear \hbox{X-ray} sources as a function of
\hbox{0.5--2.0~keV} luminosity and compare it with that observed for local
galaxies (from PC04).  We improve the source statistics available for
intermediate-redshift, off-nuclear sources by combining the multiwavelength
data within the $\approx$2~Ms \hbox{CDF-N} and $\approx$1~Ms \hbox{CDF-S} with
new \hst\ and \chandra\ observations of the Extended \chandra\ Deep Field-South
(\hbox{E-CDF-S}; Lehmer et al. 2005a).  The E-CDF-S is composed of four
contiguous $\approx$250~ks \chandra\ fields covering an $\approx$0.3 deg$^2$
region, which flanks the $\approx$1~Ms \hbox{CDF-S}.  A large fraction of the
\hbox{E-CDF-S} ($\approx$80\%) has been observed with \hst\ in two ACS filters,
$V_{606}$ and $z_{850}$, through the Galaxy Evolution from Morphology and SEDs
(GEMS; Rix et~al. 2004) survey.  The \hbox{E-CDF-S} \chandra\ observations 
can detect $z=0.1$ off-nuclear sources with projected
physical offsets of $\simgt$2~kpc and \hbox{0.5--2.0~keV} luminosities of
$\simgt 3 \times 10^{39}$ \xlum\ in the most sensitive regions. Furthermore,
$z=0.1$ sources with physical offsets of $\simgt$3~kpc and \hbox{0.5--2.0~keV}
luminosities $\simgt 3 \times 10^{40}$ \xlum\ can be detected over the entire
$\approx$0.3~deg$^2$ \hbox{E-CDF-S} field.

The Galactic column densities are $\approx$1.3 $\times$ 10$^{20}$~cm$^{-2}$ for
the \hbox{CDF-N} (Lockman 2003) and \hbox{$\approx$8.8 $\times$ 10$^{19}$}
cm$^{-2}$ for the \hbox{CDF-S} and \hbox{E-CDF-S} (Stark et al.  1992).  All of
the \hbox{X-ray} fluxes and luminosities quoted throughout this paper have been
corrected for Galactic absorption using these column densities.  Often we quote
Poisson errors with values indicating 1$\sigma$ significance levels; these
are computed following Gehrels~(1986).  $H_0$ = 70~\hbox{km s$^{-1}$
Mpc$^{-1}$}, $\Omega_{\rm m}$ = 0.3, and $\Omega_{\Lambda}$ = 0.7 are adopted
throughout this paper (Spergel et al.  2003), and the coordinates are J2000.0.

%
\section{Off-Nuclear Source Sample Construction}
%

\subsection{Sample Selection}

As discussed above, the combination of the high spatial resolution of \hst\
imaging and sensitive \hbox{X-ray} observations with \chandra\ is effective in
detecting and classifying off-nuclear \hbox{X-ray} sources out to $z \approx
0.3$.  The $\approx$2~Ms \hbox{CDF-N}, $\approx$1~Ms \hbox{CDF-S}, and
$\approx$250~ks \hbox{E-CDF-S} (hereafter CDFs) are currently the deepest
extragalactic \hbox{X-ray} surveys conducted with \chandra\ (see, e.g., Brandt
\& Hasinger 2005 for a review), and all of these surveys have good \hst\
coverage.  These observations reach \hbox{0.5--2.0~keV} sensitivity limits
ranging from 1.8 $\times 10^{-17}$ \flux\ in the \hbox{CDF-N} to 1.1 $\times
10^{-16}$ \flux\ in the \hbox{E-CDF-S}.  However, as demonstrated in $\S$~3.4.2
of Alexander et al.  (2003) and $\S$~3.3.2 of Lehmer et al. (2005a), legitimate
lower-significance \hbox{X-ray} sources below these quoted sensitivity limits
can be identified by matching such sources to associated optically bright
galaxies.  We therefore searched for off-nuclear sources using both the
published main \chandra\ catalogs of Alexander et al. (2003; \hbox{CDF-N} and
\hbox{CDF-S}) and Lehmer et al. (2005a; \hbox{E-CDF-S}) as well as additional
lower-significance \hbox{X-ray} sources detected in these fields by running the
\verb CIAO \ source-searching algorithm \verb wavdetect \ (Freeman et~al. 2002)
at a false-positive probability threshold of $1 \times 10^{-5}$.  A total of
933, 689, and 1085 \hbox{X-ray} sources in the \hbox{CDF-N}, \hbox{CDF-S}, and
\hbox{E-CDF-S}, respectively, were thus used in our searching; these numbers do
not take into account the overlapping regions of the \hbox{CDF-S} and
\hbox{E-CDF-S}.

We searched for and classified off-nuclear \hbox{X-ray} source candidates in
the CDFs using the following criteria:

\begin{enumerate}

\item An \hbox{X-ray} source is considered to be an off-nuclear candidate if
its position is offset from the optical nucleus of an optically-bright galaxy
(ACS magnitudes of $V_{606} < 21$; see below for a description of how this
limit was selected) by $\ge$1.5 $\times$ the radius of the \chandra\ positional
error circle (see equations (2) and (3) of Alexander et al.  (2003) and Lehmer
et al. (2005a), respectively) but is still observed to lie within the optical
extent of the candidate host galaxy.  The radius of the positional error circle
(\hbox{80--90\%} confidence) for \chandra\ sources detected in the CDFs ranges
from \hbox{$\approx$0\farcs3--1\farcs5} and is dependent on the \chandra\
point-spread function (PSF) size and the number of observed counts.  Our
required minimum offset of 1.5 $\times$ the radius of the \chandra\ positional
error circle was chosen empirically based on histograms showing the number of
sources with X-ray-to-optical flux ratios similar to those of active galactic
nuclei (AGNs) (i.e., sources with $\log [f_{\rm X}/f_{\rm V}] > -0.5$) versus
their X-ray-to-optical positional offsets; we found that $\approx$98\% of
AGN-like sources with $V_{606} < 24$ were matched to within 1.5 $\times$ the
radius of the \chandra\ positional error circle.  For illustrative purposes, we
provide Figure~1, which shows the logarithm of the X-ray-to-optical flux ratio
($\log [f_{\rm X}/f_{\rm V}]$) versus positional offset in units of the
\chandra\ positional error for \hbox{0.5--2.0~keV} detected sources with
optical counterparts having $V_{606} < 24$.  On the basis of our analysis, the
number of $V_{606} < 21$ nuclear sources in our sample expected to have offsets
larger than 1.5 $\times$ the radius of the \chandra\ positional error circle is
calculated to be $\approx$0.91$^{+2.11}_{-0.75}$.

\item We required that candidate host galaxies have redshifts in the range of
$0 < z \le 0.3$.  We used spectroscopic redshifts for $\approx$98\% of the
field galaxies in the CDF-N (e.g., Barger et al. 2003; Wirth et al. 2004) and
$\approx$6\% of the field galaxies in the CDF-S and E-CDF-S (e.g., Szokoly et
al. 2004).  For the remaining field galaxies in the CDF-N ($\approx$2\%) and
CDF-S and E-CDF-S ($\approx$94\%), we used high-quality photometric redshifts
from GOODS (Mobasher et al. 2004) and COMBO-17 (Wolf et al. 2004),
respectively.  We note that in the CDF-S and E-CDF-S, where the majority of the
field galaxies have only photometric redshifts via the 17-filter photometry
from COMBO-17, the redshift uncertainties are small (i.e., $\delta_z/(1+z)
\approx 0.01$).  The selected redshift range was chosen empirically by
considering the observed number of off-nuclear sources detected as a function
of redshift.  Furthermore, for redshifts larger than 0.3, the projected linear
distance corresponding to the typical \chandra\ positional error (\simgt 7~kpc)
becomes unreasonably large for detecting many off-nuclear sources.

\item In an effort to minimize confusion with unrelated background \hbox{X-ray}
sources, we further required that each off-nuclear source candidate is detected
in either the \hbox{0.5--2.0~keV} or \hbox{0.5--8.0~keV} bands and has
a \hbox{0.5--2.0~keV} luminosity $\le$10$^{41}$ \xlum.  These restrictions are
based on the observed spectral and luminosity properties of local IXOs.
\hbox{X-ray} luminosities were computed using the equation

\begin{equation}
L_{\mbox{\scriptsize{X}}} = 4 \pi d_L^2 \; f_{\mbox{\scriptsize{X}}} \; (1+z)^{\Gamma - 2}\end{equation}

\noindent where $d_L$ is the luminosity distance, $f_{\rm X}$ is the
\hbox{X-ray} flux of each source, and $\Gamma$ is the \hbox{X-ray} photon index
assuming a power-law Spectral Energy Distribution (SED).  We adopt a photon
index of $\Gamma = 1.8$ throughout this paper (see $\S$~2.2 for justification).

\end{enumerate}

The sky density of optically-bright galaxies is relatively low, and thus the
probability of finding an unrelated background \hbox{X-ray} source within the
optical extent of these galaxies is also low.  In order to estimate the total
number of background sources that may be contaminating our sample and to
determine the optimal optical-magnitude limit down to which we search for
off-nuclear sources, we followed a similar methodology to that outlined in
$\S$~2 of PC04.  We first chose a tentative optical magnitude limit down to
which we search for off-nuclear sources.  For each candidate host galaxy, we
then visually estimated the galactic area within which we would consider an
\hbox{X-ray} source to be off-nuclear.  This area was approximated by defining
an ellipse with a semimajor and semiminor axis and position angle for each
galaxy with a circle of radius 1.5 $\times$ the radius of the \chandra\
positional error circle removed.  Next, we calculated the average
\hbox{0.5--2.0~keV} sensitivity limit over the relevant area of each galaxy
using empirically calculated sensitivity maps that were calibrated
appropriately for sources detected by \verb wavdetect \ using a false-positive
probability threshold of $1 \times 10^{-5}$.  We chose to use the
\hbox{0.5--2.0~keV} band because of its low background and correspondingly high
sensitivity.  The number of expected background \hbox{0.5--2.0~keV} sources per
galaxy was estimated using the galactic area, galaxy sensitivity limit, and
best-fit number-counts relation presented in $\S$~4 of Bauer et al.  (2004).
Finally, the total number of expected contaminating background sources was
computed by summing the contributions from each galaxy.  We found that when
restricting our search to galaxies with $V_{606} < 21$, the number of expected
contaminating background sources in the entire survey is reasonably low
($\approx$2.22), so we chose \hbox{$V_{606} = 21$} as our optical magnitude
limit.  For galaxies brighter than this limit, the median (mean) number of
expected background X-ray sources was $\approx$0.0029 (0.0064) galaxy$^{-1}$.
We appropriately correct for contaminating background sources in our analyses
below.

\subsection{X-ray and Optical Properties of Off-Nuclear Sources and Host Galaxies}

Using the criteria presented above, we identified a total of 24 off-nuclear
source candidates (see Figure~2); nine of these were previously detected by
H04.  The additional source presented in H04 that is not included here (CXOCDFS
J033220.35$-$274555.3) was within 1.5~$\times$ the \chandra\ positional error
and was only detected in the \hbox{0.5--1.0~keV} band.  All of the off-nuclear
source candidates are coincident with late-type galaxies (i.e., spirals and
irregulars); this is generally consistent with studies of IXOs in the local
universe (see $\S$~1).  We also note that two of the off-nuclear sources
(CXOHDFN~J123701.99+621122.1 and CXOHDFDN~J123706.12+621711.9) are located in
galaxies having discernable bars of optical emission.  The \hbox{X-ray}
properties of the off-nuclear sources are summarized in Table~1 and additional
properties, including those of the host-galaxies, are summarized in Table~2;
median properties of are listed in the last rows of these Tables.  The median
off-nuclear source is offset by $\approx$2.1 $\times$ the \chandra\ positional
error, and only two sources (CXOCDFS~J033230.01$-$274404.0 and
CXOECDFS~J033322.97$-$273430.7) have offsets near our required minimum offset
of 1.5 $\times$ the \chandra\ positional error (see Figure~1).  CXOCDFS
J033230.01$-$274404.0 was detected and classified as being off-nuclear in both
the CDF-S and E-CDF-S data sets independently, lending additional support to
its off-nuclear classification.  

The off-nuclear \hbox{X-ray} sources span a \hbox{0.5--2.0~keV} luminosity
range of $8 \times 10^{38}$~\xlum\ to $6 \times 10^{40}$~\xlum\ and have a
median luminosity of $\approx$$8 \times 10^{39}$ \xlum.  The host galaxies have
a median redshift of \hbox{$z \approx 0.14$}, which corresponds to a lookback
time of $\approx$1.8~Gyr, and a median $V_{606}$ magnitude of 19.1.  In
Figure~3a, we show the \hbox{X-ray} luminosities of these 24 off-nuclear
sources as a function of redshift.  The fifteen new objects (filled symbols)
presented here significantly improve the source statistics at $L_{\rm
0.5-2.0~keV} \simgt 10^{39.5}$ \xlum\ and \hbox{$z = 0.15-0.3$}; many of these
sources are from the relatively wide solid-angle E-CDF-S survey.  All sources
with \hbox{0.5--2.0~keV} upper limits are detected in the \hbox{0.5--8.0~keV}
band.

We constrained the average \hbox{X-ray} spectral shape of our off-nuclear
sources by stacking the \hbox{0.5--2.0~keV} and \hbox{2--8~keV} source counts
and exposures.  We then computed a vignetting-corrected band ratio by taking
the ratio $\Phi_{\rm 2-8~keV}/\Phi_{\rm 0.5-2.0~keV}$, where $\Phi$ is defined
to be the \hbox{X-ray} count rate in units of counts~s$^{-1}$.  The band ratio
was converted into a mean effective photon index $\Gamma_{\rm eff}$ using the
\chandra\ \hbox{X-ray} Center's Portable, Interactive, Multi-Mission Simulator
(PIMMS).  For the 24 off-nuclear \hbox{X-ray} sources, the mean effective
photon index is \hbox{$\Gamma_{\rm eff} = 1.83_{-0.17}^{+0.21}$}, a value
consistent with that for IXOs observed in the local universe (e.g., Liu \&
Mirabel 2005 and references therein).  We therefore adopted $\Gamma = 1.8$ when
computing \hbox{X-ray} luminosities for our off-nuclear sources using
equation~(1).

Since all of our off-nuclear sources are coincident with late-type galaxies
(see Figure~2), we chose to restrict further comparative analyses to galaxies
with late-type morphologies.  We selected galaxies in the ACS-covered regions
of the CDFs and required $V_{606} < 21$ in the redshift range $0 < z \le 0.3$.
At the quoted optical-magnitude limit, roughly half of the objects detected are
Galactic stars; these were characterized visually as point-like sources with
diffraction spikes and removed from our list of candidate field galaxies.  In
total, 385 objects were selected as being non-stellar extragalactic sources
with $V_{606} < 21$ and \hbox{$0 < z \le 0.3$}.  We visually classified each
galaxy as being either an elliptical (48 galaxies) or a spiral or irregular (337
galaxies).  Figure~3b shows the estimated rest-frame 6000 \AA\ luminosities
(\hbox{$\nu L_{\nu}$ [6000~\AA]}) for all of the 337 spiral and irregular galaxies
in our sample (small filled circles).  Galaxies hosting off-nuclear sources are
plotted with larger open and filled symbols (for H04 and new sources,
respectively); most of these occupy the high-luminosity end of the
distribution.  Optical luminosities were estimated using the $V_{606}$
magnitude and applying a $K$-correction assuming an Scd spiral-galaxy optical
SED (Coleman et~al.  1980); this SED was chosen since it is commonly observed
for star-forming galaxies.  We chose to use the $V_{606}$ band to compute the
6000 \AA\ luminosities because (1) it is availabile over all of the CDFs and
(2) the 6000 \AA\ continuum traces the emission from relatively old stellar
populations.  The $K$-corrections for the galaxies are small ($\simlt 0.4$~mag)
and have little dependence on galaxy SED choice.  The median 6000 \AA\
luminosities (6000 \AA\ absolute magnitudes, $M_{606}$) of the
intermediate-redshift field galaxies and the host galaxies of off-nuclear
\hbox{X-ray} sources are \hbox{$\nu L_{\nu}$ (6000~\AA)~$\approx$~1.9 $\times
10^{43}$~\xlum} ($-$19.8) and $\approx$2.7 $\times 10^{43}$~\xlum\ ($-$20.2),
respectively.

Ten of our 24 (41$^{+18}_{-13}$\%) off-nuclear \hbox{X-ray} sources appear to
be coincident with optical knots of emission (see column~8 of Table~2 and the
images in Figure~2).  In the local universe, IXOs often appear to be located
within or near star-forming regions, ranging from small ($< 100$~pc diameter)
diffuse H$\alpha$ emission complexes to giant ($\simgt$500~pc diameter)
H~{\small II} regions (e.g., Pakull \& Mirioni 2002; Liu \& Bregman 2005;
Ramsey et al. 2006).  The optical knots in our sample have apparent diameters
of $\approx$500--1000~pc and optical luminosities of \hbox{$\nu L_{\nu}$
(6000~\AA) $\approx 10^{40-41}$~\xlum}.  These optical properties are broadly
consistent with those of giant H~{\small II} regions in the local universe
(e.g., Kennicutt~1984), suggesting these off-nuclear sources trace distant
star-formation regions.  In the GOODS fields, high-resolution ACS imaging is
available over four optical bands ($B_{435}$, $V_{606}$, $i_{775}$, and
$z_{850}$); for the eight off-nuclear sources with optical knots located in
these ACS regions, we analyzed the colors in the immediate vicinity of the knots.
Figure~4 shows the relative color difference between the optical knots and
their host galaxies.  These relative colors are defined as follows:

$$\Delta (B_{435}-V_{606}) \equiv (B_{435}-V_{606})_{\rm knot} - (B_{435}-V_{606})_{\rm galaxy}$$
\begin{equation}
\Delta (V_{606}-i_{775}) \equiv (V_{606}-i_{775})_{\rm knot} - (V_{606}-i_{775})_{\rm galaxy} \rm{.}
\end{equation}

\noindent We find that the optical knots are relatively blue compared to their
host galaxies (i.e., mean values of $\Delta (B_{435}-V_{606}) = -0.22 \pm 0.14$ and
$\Delta (V_{606}-i_{775})=-0.13 \pm 0.18$ are observed).  We stacked the
\hbox{0.5--2.0~keV} and \hbox{2--8~keV} counts from our off-nuclear sources
with and without optical knots to see if we could distinguish between the mean
effective photon indices ($\Gamma_{\rm eff}$; see text above) of these two
populations.  We found that the average spectral shapes of these two
populations are statistically consistent (i.e., $\Gamma_{\rm
eff}=1.90^{+0.26}_{-0.21}$ and $1.70^{+0.41}_{-0.28}$, for sources with and
without optical knots, respectively).

We also investigated the optical colors of the host galaxies to see if the
off-nuclear sources are found in more actively star-forming galaxies.  We
utilized $B$- and $V$-band magnitudes from ground-based observations of the
\hbox{CDF-N} with the Subaru 8.2 m telescope (Capak et al. 2004), and of the
\hbox{CDF-S} and \hbox{E-CDF-S} with the Wide Field Imager of the MPG/ESO
telescope at La Silla (see $\S$~2 of Giavalisco et al. 2004); small corrections
for the differing bandpasses were applied to the \hbox{CDF-S} and
\hbox{E-CDF-S} magnitudes to match those in the \hbox{CDF-N}.  Optical emission
lines produced by star-forming galaxies can influence the observed broad-band
fluxes.  At the median redshift of our off-nuclear source sample ($z=0.14$),
the strong nebular emission line due to [O~{\small II}] $\lambda$3727 is
located in the $B$-band and the redder emission lines from the [O~{\small III}]
$\lambda \lambda$4959, 5007 doublet and H$\beta$ $\lambda$4861 are located in
the $V$-band.  In Figure~5, we show the $B-V$ colors vs. $V_{606}$-band
magnitudes for our 337 spiral and irregular galaxies in the CDFs.  Field
galaxies hosting off-nuclear sources have been outlined with open squares and
circles, which distinguish between sources with and without optical knots,
respectively.  The mean $B-V$ optical colors have been plotted for the galaxies
(solid line), galaxies hosting off-nuclear sources (dot-dashed line), and
galaxies hosting off-nuclear sources with (dashed line) and without (dotted
line) optical knots.  The typical galaxy hosting an off-nuclear \hbox{X-ray}
source shows somewhat bluer optical colors than typical galaxies in the field.
We utilized the two-sample Kolmogorov-Smirnov (K-S) test on the unbinned $B-V$
samples of field galaxies and galaxies hosting off-nuclear sources, and
repeated the test to compare the $B-V$ sample of field galaxies with
off-nuclear sources with and without optical-knot counterparts.  We found that,
at the $\approx$89\% confidence level (i.e., K-S probability~$\approx$~0.11),
the $B-V$ colors of galaxies hosting off-nuclear sources are statistically
different from typical field galaxies in the CDFs.  Furthermore, the colors of
galaxies hosting off-nuclear sources with optical knots are statistically
different from those of typical field galaxies at the $\approx$96\% confidence
level, but galaxies hosting off-nuclear sources without optical knots have
colors statistically consistent with those of typical field galaxies.  These
tests suggest that off-nuclear sources are indeed preferentially located in
galaxies undergoing intense star formation; the optical knots directly trace
star-forming complexes in such galaxies.  For comparison, we plotted the
distribution of Johnson $U-B$ colors versus $V$-band magnitude for RC3 galaxies
(inset to Figure~5) and highlighted six local galaxies (NGC~1566, NGC~1672,
NGC~3623, NGC~4088, NGC~4303, and NGC~4490) observed to host IXOs coincident
with distinct, luminous optical knots from Liu \& Bregman~(2005).  These
galaxies appear to be notably bluer than typical RC3 galaxies, and using the
K-S test, we find that the colors of these galaxy samples are statistically
different at the $\approx$96\% confidence level.  Locally, the $U$- and
$B$-band magnitudes sample the strong emission features expected in the $B$-
and $V$-bands, respectively, at \hbox{$z =0.14$}.

%
\section{Analysis and Results}
%

A primary goal of this investigation is to determine whether the ``true''
luminosity-dependent fraction of galaxies containing off-nuclear \hbox{X-ray}
sources ($f_{\rm T}$) evolves with cosmic time (see $\S$~1).  Below, we
describe our procedure for computing this fraction for our
intermediate-redshift galaxy sample and a matched comparison sample of local
galaxies from PC04.  We assess observational constraints on the \hbox{X-ray}
luminosity detection limit and angular resolution for both our sample and that
of PC04.  To this end, we first describe the computation of the
luminosity-dependent ``observed'' fraction of galaxies hosting off-nuclear
sources ($f_{\rm O}$); this takes into account the spatially varying
sensitivity of the \chandra\ observations.  We then estimate the true fraction
($f_{\rm T}$) by calculating correction factors to $f_{\rm O}$, which account
for (1) the number of off-nuclear sources we expect to miss due to
angular-resolution limitations (i.e., the number of off-nuclear sources with
offsets smaller than the resolution limit) and (2) the multiplicity of these
sources within host galaxies (i.e., the expected number of off-nuclear sources
within a galaxy having at least one off-nuclear source). 

\subsection{The Observed Fraction ($f_{\rm O}$)}

To compute the observed fraction of intermediate-redshift galaxies hosting
off-nuclear \hbox{X-ray} sources, we followed a procedure similar to that
outlined in $\S$~2 of PC04.  For each of our 337 spiral and irregular galaxies, we
used the \hbox{0.5--2.0~keV} sensitivity maps described in $\S$~2.1 and
corresponding redshift information to compute an \hbox{X-ray} luminosity limit
above which we could detect an off-nuclear source.  All luminosities were
calculated using equation~(1), adopting $\Gamma=1.8$.  In Figure~6a, we show
the number of galaxies with \hbox{X-ray} coverage sensitive enough to detect an
off-nuclear source of a given \hbox{0.5--2.0~keV} luminosity, \Lx.  For
example, there are 20 galaxies with \hbox{X-ray} coverage sensitive enough to
detect an off-nuclear source with \hbox{\Lx\ $\approx 10^{38.9}$ \xlum} (i.e.,
the leftmost bin of Figure~6a), and there are 337 galaxies with coverage
sensitive enough to detect an off-nuclear source with \Lx\ $\approx 10^{41}$
\xlum\ (i.e., the rightmost bin of Figure~6a).  Figure~6b shows the number of
galaxies in each \Lx\ bin of Figure~6a observed to host an off-nuclear source
of \Lx\ or greater.  In Figure~6b, the number of galaxies included in a given
\Lx\ bin versus its neighboring lower-luminosity \Lx\ bin is affected by both
the addition of galaxies hosting off-nuclear sources with less sensitive X-ray
coverage and the subtraction of galaxies hosting off-nuclear sources that fall
below \Lx.  In order to aid in the understanding of this progression, we have
created Figures~6c and 6d, which show the number of added and subtracted
galaxies (in ascending order from \Lx $\approx 10^{38.9}$ \xlum\ to \Lx
$\approx 10^{41}$ \xlum), respectively.  For example, six galaxies in the first
\Lx\ bin of Figure~6a that host off-nuclear sources with \hbox{\Lx\ $\simgt
10^{38.9}$ \xlum} are added to the first \Lx\ $\approx 10^{38.9}$ \xlum\ bin of
Figure~6b.  In the next higher \Lx\ bin (i.e., \Lx\ $\approx 10^{39.3}$ \xlum),
two new galaxies hosting off-nuclear sources with \Lx\ $\simgt 10^{39.3}$
\xlum\ are added to the appropriate \Lx\ bin of Figure~6b, and two galaxies
with off-nuclear source luminosities $< 10^{39.3}$ \xlum\ are subtracted from
the same bin.  Thus the total number of galaxies with off-nuclear sources of
\Lx\ $\ge 10^{39.3}$ \xlum\ remains at six.  

In order to calculate the observed fraction of intermediate-redshift galaxies
hosting off-nuclear sources ($f_{\rm O}$), we divide the histogram entries of
Figure~6b, with the estimated number of background sources subtracted (see
$\S$~2.1), by those of Figure~6a; the observed fraction is presented in
Figure~7 as a dashed line with shaded 1$\sigma$ error envelope (computed
following Gehrels~1986).  The dashed line terminates for \Lx~$\simgt
10^{40.7}$~\xlum\ due to the fact that there are no off-nuclear sources in our
sample with luminosities exceeding $10^{40.7}$~\xlum; the shaded region
therefore represents the 3$\sigma$ upper limit to $f_{\rm O}$ for \Lx~$\simgt
10^{40.7}$~\xlum.  To make comparisons between the observed fraction of our
intermediate-redshift sample and that of the local sample of PC04, we matched
the optical-luminosity distribution of the PC04 spiral and irregular galaxies
(i.e., Hubble type $\ge$~0) to that of our 337 spiral and irregular galaxies.
At the median redshift ($z = 0.14$) of our off-nuclear \hbox{X-ray} source
sample, the $V_{606} < 21$ requirement on the inclusion of galaxies in our
analyses equates to requiring the rest-frame optical luminosity of a particular
galaxy to be $\nu L_{\nu}$(6000 \AA) $\simgt 10^{42.6}$ \xlum\ ($M_{606} \simlt
-18.1$; see Figure~3b).  We created a matched PC04 subsample by applying the
same optical luminosity cut to the original PC04 spiral and irregular sample.
In total, 329 spiral and irregular galaxies were selected for comparison from
the original PC04 sample of 766 galaxies.  Figure~8 shows the $\nu
L_{\nu}$(6000 \AA) luminosity distributions of galaxies for both our sample
(solid histogram) and the matched subsample of PC04 (dotted histogram).  The
two luminosity distributions span roughly the same $\nu L_{\nu}$(6000 \AA)
range and have similar overall shapes; a K-S test indicates that these two
distributions are statistically consistent for $\nu L_{\nu}$(6000 \AA) $\simgt
10^{42.6}$ \xlum\ (i.e., K-S probability $\approx$~0.6).  Using the matched PC04
subsample and the procedure described above, we recomputed the relevant
observed fraction of local galaxies hosting IXOs ($f_{\rm O,PC04}$); this is
shown in Figure~7 as filled circles with 1$\sigma$ error bars and 3$\sigma$
upper limits.  

\subsection{The True Fraction ($f_{\rm T}$)}

As noted above, the angular resolution of \chandra, while superb, is limited to
the classification of off-nuclear sources at intermediate-redshifts that are
offset by more than 1.5 $\times$ the \chandra\ positional error from the
optical position of the galactic nucleus.  Furthermore, the classification of
IXOs in the PC04 sample is also limited to \hbox{X-ray} sources offset by
$\simgt$10\arcsec\ (Colbert \& Ptak 2002).  Therefore, the observed fractions
for both our sample and the matched PC04 subsample can only be considered to be
lower limits on the true fractions of galaxies hosting off-nuclear sources, and
therefore direct comparison between these two samples is not meaningful.  To
address this problem, we have computed the relevant scaling between the
observed and true fractions (for both our sample and the matched PC04
subsample) through simulations using a sample of local IXO-hosting galaxies
observed with \chandra; we describe these simulations below.   

We calculated the distribution of projected physical offsets of 65 IXOs from 29
\chandra-observed local ($D \simlt 50$~Mpc) galaxies (compiled by Liu \&
Mirabel 2005; see references therein).  Observations of IXOs in local galaxies
with \chandra\ should be statistically complete for sources with offsets
$\simgt$0.1~kpc, assuming a positional accuracy of $\approx$0\farcs5 and a
maximal galactic distance of $\approx$50~Mpc.  The normalized distribution of
offsets for the 65 IXOs used here is shown in Figure~9 (solid histogram).
Similar distributions were calculated using subsets of these IXOs with
\hbox{0.5--2.0~keV} luminosities $> 10^{39-39.5}$ \xlum; no strong luminosity
dependence was observed in the shape of these curves.  The overall offset
distribution shown in Figure~9 indicates that IXOs are often located relatively
close to their host-galaxy nucleus, which is considerably different from the
scenario where IXOs are distributed throughout galaxies with constant density
(dotted curve).  This observed distribution is likely due to the high
circumnuclear star-formation activity commonly found in normal and starburst
galaxies (e.g., Kennicutt~1998) and is consistent with the IXO offset
distribution noted by Swartz et~al. (2004).  The median physical offset for the
65 \chandra-observed local IXOs is $\approx$3~kpc (vertical dashed line) and
the expected resolution for CDF sources at the median redshift ($z=0.14$) of
our off-nuclear source sample is $\approx$2.6~kpc (vertical dot-dashed line);
to first order, this suggests that our observed off-nuclear sources only make
up $\sim$50\% of the physical offset distribution, and we are plausibly missing
about half of the off-nuclear sources.

Next, we computed the fraction of galaxies in which an \hbox{X-ray} source of
a given linear offset would be classified as ``off-nuclear'' for galaxies in
both our intermediate-redshift galaxy sample and the matched PC04 subsample.
The relevant fractions for each sample are shown as solid curves in the top
panels of Figure~10; dotted curves show the case where there is no
resolution limitation.  These offset-dependent fractions are constrained by
both the projected physical resolution (the observed rise in the fraction near
small offsets) and the projected optical size of the galaxies (the observed
decline in the fraction at larger offsets).  The angular resolution was taken
to be 1.5 $\times$ the \chandra\ positional error for our galaxy sample and was
fixed at 10\arcsec\ for the matched PC04 subsample.  The angular size of each
galaxy was assumed to be the apparent optical semimajor axis (see $\S$~2.1) and
0.5 $\times$ the RC3 major-axis diameter ($D_{25}$) for our sample and the
matched PC04 subsample, respectively.  The solid curves presented in the top
panels of Figure~10 for each sample are significantly different in the
small-offset regime, showing that our sample is more heavily affected by
angular-resolution limitations than the matched PC04 subsample.  We simulated
the expected observable offset distribution of each galaxy sample by convolving
the local offset distribution presented in Figure~9 (solid histogram) with the
detectable fraction curves (top panels of Figure~10; solid curves); these are
displayed as solid histograms in the bottom panels of Figure~10.  For
comparison, we have shown the actual observed offset distributions of
off-nuclear sources in our sample and the matched PC04 subsample (dashed
histograms in the bottom panels of Figure~10, respectively).  By inspection, we
find that the observed and simulated distributions are consistent.

In order to estimate the number of off-nuclear sources that were missed due to
instrumental-resolution limitations, we simulated the expected distributions of
offsets for off-nuclear sources in each galaxy sample for the case where there
is no resolution limitation.  This was done by convolving the local offset
distribution function (solid histogram in Figure~9) with the dotted curves in
the top panels of Figure~10.  These resulting distribution functions, normalized
by the simulated observed distribution functions (solid histograms in the
bottom panels of Figure~10), are presented as dotted histograms in the bottom
panels of Figure~10.  These calculations suggest that 62.1\% and
35.0\% of the off-nuclear sources would remain unclassified as
``off-nuclear'' for our intermediate-redshift sample and the matched PC04
subsample, respectively.  

In the simple case where we would expect $\approx$1 off-nuclear source per galaxy, we
could simply rescale the observed fractions by a constant scaling factor to
obtain the true fractions; we define this scaling factor as $\alpha_{\rm s}$,
which is 2.64 and 1.54 for our off-nuclear source sample and the matched PC04
subsample, respectively.  However, observations of IXOs in the local universe,
show that, on average, there is more than one IXO per IXO-hosting galaxy (e.g.,
Colbert \& Ptak~2002).  Since we observe only one IXO per IXO-hosting galaxy
for our intermediate-redshift sample, a simple constant scaling of the observed
fraction to obtain the true fraction is not appropriate.  Hereafter, we refer
to the mean number of observed IXOs per IXO-hosting galaxy as the
``multiplicity'' factor ($m$), which varies with IXO luminosity; using the Liu
\& Mirabel~(2005) sample of IXOs, we have estimated the ``true''
luminosity-dependent multiplicity factor ($m_{\rm T}$).  We find that $m_{\rm
T}$ has a value of $\approx$2.1 and $\approx$1.0 IXOs per IXO-hosting galaxy
for \hbox{0.5--2.0~keV} luminosities of \Lx~$\simgt 10^{38.9}$~\xlum\ and
\Lx~$\simgt 10^{40}$~\xlum, respectively; for comparison, the corresponding
observed PC04 multiplicity factor ($m_{\rm O,PC04}$) has a value of
$\approx$1.5 and $\approx$1.0 IXOs per IXO-hosting galaxy, respectively.  Using
the scaling factor $\alpha_{\rm s}$, the observed multiplicity factor $m_{\rm
O}$, and the true multiplicity factor $m_{\rm T}$, we converted the observed
fraction $f_{\rm O}$ to the true fraction $f_{\rm T}$ following

\begin{equation}
f_{\rm T}=\alpha_{\rm s}\frac{m_{\rm O}}{m_{\rm T}}f_{\rm O} \rm{.}
\end{equation}

\noindent The luminosity-dependent scaling factors (for our
intermediate-redshift sample and the matched PC04 subsample), $\alpha_{\rm
s}m_{\rm O}/m_{\rm T}$, which scale $f_{\rm O}$ to obtain $f_{\rm T}$, are
shown in Figure~11a; the resulting true fractions with 1$\sigma$ errors
(computed following Gehrels 1986) are shown in Figure~11b.  Fractions measured
for our sample ($f_{\rm T,int-{\it z}}$) are shown as the dashed line
surrounded by the shaded error envelope, and the matched PC04 fractions
($f_{\rm T,PC04}$) are plotted as filled black circles with 1$\sigma$ error
bars and 3$\sigma$ upper limits.  The dotted curve shows the fraction of
our galaxies with an \hbox{X-ray} source of \hbox{0.5--2.0~keV} luminosity \Lx\ or
greater (including both nuclear and off-nuclear sources).  This curve was
obtained by matching galaxies in our sample to \hbox{X-ray} sources from the
main and supplementary \chandra\ catalogs of the CDFs using a matching criterion
of less than one semimajor axis length.  Matched sources may include a variety
of \hbox{X-ray} sources such as AGNs, luminous nuclear starbursts, and
off-nuclear \hbox{X-ray} sources with both small and large offsets (i.e.,
offsets both less than and greater than 1.5 $\times$ the radius of the \chandra\ positional
error circle).  This curve represents an upper limit to the true fraction of
galaxies hosting off-nuclear sources, and we note that our calculated true
fraction is below this limit.  Furthermore,
based on this curve, we calculate that off-nuclear sources are found in
$\approx$75\% and $\approx$40\% of the $z \le 0.3$ galaxies detected in the
\hbox{0.5--2.0~keV} bandpass with coverage sensitive enough to detect sources
of \Lx~$> 10^{39}$ \xlum\ and \Lx~$> 10^{40.7}$ \xlum, respectively.  
For further comparison, we have plotted the fraction of
galaxies with an \hbox{X-ray} source of \hbox{0.5--2.0~keV} luminosity \Lx\ or
greater but with a luminosity upper bound of $10^{41.5}$~\xlum\ (dot-dashed
curve); this upper limit was adopted to include mostly galaxies powered by
star-forming processes.  Again, our estimate of $f_{\rm T,int-{\it z}}$ appears
to be reasonable in comparison to this detection fraction.

Figure~11b shows suggestively that the off-nuclear source frequency for field
galaxies rises with redshift.  We estimate that $\approx$31$^{+20}_{-16}$\% of
intermediate-redshift spiral and irregular galaxies with $\nu L_{\nu}$(6000
\AA) $\simgt 10^{42.6}$ \xlum\ host off-nuclear sources with
\Lx~$\simgt$~10$^{39}$~\xlum\ versus $\approx$16$_{-4}^{+5}$\% in the local
universe (errors are 1$\sigma$).  As mentioned in $\S$~1, one may plausibly
expect that the frequency of off-nuclear sources would rise as a function of
redshift due to the observed global increase in star-formation density, which
is measured to be $\approx$1.2--3.0 times higher at $z \approx 0.05-0.3$ than
it is in the local universe (e.g., P{\'e}rez-Gonz{\'a}lez et~al. 2005;
Schiminovich et al. 2005).  Furthermore, since the number of IXOs in spiral and
irregular galaxies is observed to increase linearly with \hbox{star-formation}
rate (e.g., Swartz et al.  2004), it is reasonable to expect that the frequency
of off-nuclear source incidence for field galaxies would roughly scale linearly
with the \hbox{star-formation} density.  In Figure~12 we show the ratio of
off-nuclear source incidence fraction for our intermediate-redshift sample and
the matched PC04 subsample (i.e., $f_{\rm T,int-{\it z}}$/$f_{\rm T,PC04}$) as
filled circles with 1$\sigma$ error bars.  Errors on this quantity were
computed using the error propogation methodology outlined in $\S$~1.7.3 of
Lyons (1991).  The fraction ratio for off-nuclear sources with
\Lx~$\simgt$~10$^{39}$~\xlum\ is $\approx$1.9$^{+1.4}_{-1.1}$; this is elevated
from unity at the $\approx$80\% confidence level.  The dashed horizontal line
shows the median fraction ratio.  The shaded region shows the expected ratios
for the case where the off-nuclear source incidence fraction scales with
star-formation density; the dotted horizontal line shows the case where there
is no evolution.  We note that these computed ratios appear to be broadly
consistent with the expected scaling of off-nuclear source incidence with
redshift due to the increased global star-formation density.

\subsection{Consistency Check}

We have performed consistency checks on the results above by degrading the
resolution of the PC04 subsample to match that of our intermediate-redshift
sample and doing similar analyses to those above.  This was achieved by
calculating the median physical resolution of our intermediate-redshift galaxy
sample (i.e., the median physical offset corresponding to 1.5 $\times$ the
\chandra\ positional error) and generating a subsample of PC04 IXOs with
offsets greater than this median offset.  Using this subset of IXOs, we
calculated the resolution-normalized observed fraction $f_{\rm O,PC04}^{\rm
norm}$.  In comparison to the intermediate-redshift observed fraction $f_{\rm
O,int-z}$, we find results consistent with those presented above.  For example,
we find that for sources with \Lx~$\simgt$~10$^{39}$~\xlum, the
resolution-normalized fraction ratio $f_{\rm O,int-{\it z}}$/$f_{\rm
O,PC04}^{\rm norm} \approx 2.7^{+2.0}_{-1.3}$.

%
\section{Summary and Future Work}
%

We have presented the largest sample to date of intermediate-redshift ($z
\simlt 0.3$; \hbox{$z_{\rm median}=0.14$}), off-nuclear \hbox{X-ray} sources
hosted in optically-bright ($V_{606} < 21$) field galaxies in the \chandra\
deep fields.  These off-nuclear \hbox{X-ray} sources were found to have similar
\hbox{X-ray} spectral shapes and optical environments to IXOs in the local
universe and are exclusively found to be coincident with late-type spiral and
irregular galaxies.  Using this sample, we found that the fraction of spiral
and irregular galaxies hosting an off-nuclear \hbox{X-ray} source as a function
of \hbox{X-ray} luminosity is suggestively higher at intermediate-redshifts;
for off-nuclear sources with \hbox{0.5--2.0~keV} luminosities
$\simgt$~10$^{39}$~\xlum, this fraction is measured to be
$\approx$31$^{+20}_{-16}$\% for intermediate-redshift field galaxies versus
$\approx$16$_{-4}^{+5}$\% for local galaxies (see Figure~11).  In computing
this fraction, we have accounted for the facts that (1) the \hbox{X-ray}
sensitivity limit varies spatially over these fields and (2) the angular
resolution of \chandra\ limits the classification of off-nuclear sources.

Although the angular-resolution limitations of \chandra\ would remain, the
current situation could still be improved by future \chandra\ observations over
these fields.  We note that only $\approx$20 of the spiral and irregular
galaxies ($\approx$6\%) in our sample have \hbox{X-ray} coverage sensitive
enough to detect off-nuclear sources of \hbox{0.5--2.0~keV} luminosity
\Lx~$\approx 10^{39}$~\xlum\ (see Figure~6a), a regime where $\approx$31\% of
the galaxies are expected to host off-nuclear \hbox{X-ray} sources; these
sources are dispersed over all of the CDFs.  Deeper \chandra\ observations over
the CDFs would not only improve the source statistics in this regime but
would also improve the positional accuracies of brighter sources too.  In the
$\approx$2~Ms CDF-N, sources near the aim point that have sufficient photon
counts have positional accuracies of $\approx$0\farcs3.  At this resolution,
field galaxies at $z = 0.1$ and $z = 0.3$ with off-nuclear sources offset by
$\simgt$0.8~kpc and $\simgt$2.0~kpc, respectively, could be identified, and
more stringent constraints could be placed on the statistical properties of
these sources.  Such observations covering the presently-classified off-nuclear
sources would improve our knowledge of their spectral and variability
properties.  Furthermore, we note that other programs that utilize deep
\chandra\ exposures in combination with \hst\ observations (e.g., the Extended
Groth Strip; Nandra et al. 2005) could build upon the present
intermediate-redshift, off-nuclear source sample and improve the statistical
constraints on their frequency in field galaxies.

Presently, \chandra\ is the only observatory capable of classifying and
characterizing intermediate-redshift, off-nuclear \hbox{X-ray} sources.
\hbox{X-ray} missions of the relatively near future (e.g., \conx\ and
\xeus)\footnote{For further information regarding \conx\ and \xeus, visit
http://constellation.gsfc.nasa.gov/ and http://www.rssd.esa.int/XEUS/,
respectively.} will be capable of placing tighter constraints on the spectral
and temporal properties of these sources.  
However, our understanding of the statistical properties of this population and
its evolution with cosmic time can only be substantially improved by future
\hbox{X-ray} missions with subarcsecond imaging capabilities such as
\hbox{\genx},\footnote{For further information regarding \genx, visit
http://genx.cfa.harvard.edu.} which is planned to have imaging
($\approx$0\farcs1 resolution) and sensitivity capabilities which greatly
supersede those already available through \chandra.  At $z \approx 2$, an
off-nuclear source with an \hbox{X-ray} luminosity of $\approx$$10^{39}$~\xlum\
and a physical offset of $\simgt$0.8~kpc could be detected and classified in a
moderate-length \hbox{\genx} exposure.  Somewhat more luminous off-nuclear
sources (\Lx~$\simgt 10^{39.5}$~\xlum) with offsets as small as
$\approx$0.6~kpc could be isolated using \hbox{\genx} at $z \approx 4$.  

Our results suggest that the majority of the \hbox{X-ray} activity from normal
and starburst galaxies can be explained by the presence of luminous
(\Lx~$\simgt 10^{39}$~\xlum) off-nuclear \hbox{X-ray} sources (see, e.g.,
Figure~11b); this is consistent with studies of \hbox{X-ray} point-source
populations in local galaxies (e.g., Colbert et al. 2004).  As discussed in
$\S$~1, the global star formation rate increases with redshift, and the mean
\hbox{X-ray} luminosity for field galaxies also rises as a function of
redshift.  If this increase in the \hbox{X-ray} emission is indeed mainly due
to the presence of many luminous off-nuclear sources associated with star
formation, then we would expect that for distant, energetic star-forming
galaxies such as Lyman break galaxies (LBGs), there may be significant crowding
of luminous off-nuclear sources.  The mean \hbox{0.5--2.0~keV} luminosity for
LBGs at $z \approx 3$ is $\approx$10$^{41}$~\xlum\ (e.g., Lehmer et al. 2005b).
If we assume that the majority of this emission is coming from luminous
off-nuclear sources, then we may expect that the average LBG would have
$\approx$5--10 off-nuclear sources each with \Lx~$\approx$~10$^{39.9}$~\xlum.
When adopting an angular radius of $\approx$0\farcs3 for a typical $z \approx
3$ LBG (e.g., Ferguson et al. 2004), we find that the corresponding off-nuclear
source density and mean angular separation would be
$\approx$20--40~arcsec$^{-2}$ and $\approx$0\farcs16--0\farcs22, respectively.
Future missions such as \hbox{\genx} would suffer from non-negligible source
confusion from the multiple off-nuclear sources within these distant LBGs.

\acknowledgements

We thank Andrew Ptak for sharing data and Mike Eracleous, Caryl Gronwall, and
Ohad Shemmer for useful suggestions and discussions.  We thank the anonymous
referee for detailed comments and suggestions, which have improved this
manuscript.  We gratefully acknowledge the financial support of NSF CAREER
award AST-9983783 (B.D.L., W.N.B.), \chandra\ X-ray Center grant G04-5157A
(B.D.L., W.N.B., A.T.S.), the Royal Society (D.M.A.), the \chandra\ Fellowship
program (F.E.B.), and NSF grant AST 03-07582 (D.P.S.).

\clearpage

%
{} 
%

%
%

\begin{rotate}

\begin{deluxetable}{lcccccccc}
\tablenum{1}
\tabletypesize{\scriptsize}
\tablewidth{0pt}
\tablecaption{Off-Nuclear Sources: X-ray Properties}

\tablehead{
\colhead{}       &
\multicolumn{2}{c}{X-ray Counts}             &
\colhead{}       &
\colhead{$f_{\rm 0.5-2.0~keV}$}        &
\colhead{$f_{\rm 0.5-8.0~keV}$}            &
\colhead{$L_{\rm 0.5-2.0~keV}$}   &
\colhead{$L_{\rm 0.5-8.0~keV}$}       &
\colhead{}                         \\

\colhead{Source Name}                   &
\colhead{0.5--2.0~keV}       &
\colhead{0.5--8.0~keV}       &
\colhead{$\Gamma$}                      &
\colhead{($10^{-16}$ [cgs])}       &
\colhead{($10^{-16}$ [cgs])}       &
\colhead{($\log$ cgs)}                    &
\colhead{($\log$ cgs)}                    &
\colhead{Survey}                   \\

\colhead{(1)}         &
\colhead{(2)}         &
\colhead{(3)}         &
\colhead{(4)}         &
\colhead{(5)}         &
\colhead{(6)}         &
\colhead{(7)}         &
\colhead{(8)}         &
\colhead{(9)}         
}

\startdata
CXOECDFS J033122.00$-$273620.1 &                             $<$11.5 &                17.0$^{+5.8}_{-5.8}$ &        1.8$^{\rm a}$ &  $<$3.36 &     8.37 &           $<$40.2 &              40.6 & E-CDF-S 02 \\
CXOECDFS J033128.84$-$275904.8 &                10.9$^{+4.8}_{-3.6}$ &                             $<$13.4 &        1.8$^{\rm a}$ &     2.78 &  $<$5.71 &              40.8 &           $<$41.1 & E-CDF-S 03 \\
CXOECDFS J033139.05$-$280221.1 &                 1.5$^{+1.4}_{-1.4}$ &                             $<$14.6 &        1.8$^{\rm a}$ &     0.43 &  $<$7.03 &              39.9 &           $<$41.1 & E-CDF-S 03 \\
CXOECDFS J033143.46$-$275527.8 &                              $<$4.7 &                 5.4$^{+2.5}_{-2.5}$ &        1.8$^{\rm a}$ &  $<$1.18 &     2.26 &           $<$39.6 &              39.9 & E-CDF-S 03 \\
CXOECDFS J033143.48$-$275103.0 &                10.2$^{+5.1}_{-3.8}$ &                             $<$16.3 &        1.8$^{\rm a}$ &     2.76 &  $<$7.41 &              40.8 &           $<$41.2 & E-CDF-S 03 \\
 CXOCDFS J033219.10$-$274445.6 &                 5.0$^{+2.4}_{-2.4}$ &                             $<$15.6 &        1.8$^{\rm a}$ &     0.33 &  $<$1.74 &              39.6 &           $<$40.3 &      CDF-S \\
\\
 CXOCDFS J033221.91$-$275427.2 &                 6.9$^{+3.2}_{-3.2}$ &                             $<$21.5 &        1.8$^{\rm a}$ &     0.46 &  $<$2.43 &              39.6 &           $<$40.3 &      CDF-S \\
 CXOCDFS J033230.01$-$274404.0 &              87.6$^{+11.2}_{-10.1}$ &             104.0$^{+12.8}_{-11.6}$ &                  1.9 &     6.39 &    12.28 &              40.0 &              40.2 &      CDF-S \\
 CXOCDFS J033234.73$-$275533.8 &                             $<$20.6 &              58.6$^{+12.9}_{-11.8}$ &        1.8$^{\rm a}$ &  $<$1.48 &     7.05 &           $<$38.7 &              39.4 &      CDF-S \\
CXOECDFS J033249.26$-$273610.6 &                38.6$^{+7.8}_{-6.6}$ &                56.7$^{+9.7}_{-8.4}$ &                  1.6 &     9.78 &    27.46 &              40.5 &              40.9 & E-CDF-S 01 \\
CXOECDFS J033316.29$-$275040.7 &                22.1$^{+6.7}_{-5.6}$ &                             $<$22.1 &        1.8$^{\rm a}$ &     5.90 &  $<$9.95 &              40.1 &           $<$40.3 & E-CDF-S 04 \\
CXOECDFS J033322.97$-$273430.7 &                 5.3$^{+2.5}_{-2.5}$ &                             $<$15.1 &        1.8$^{\rm a}$ &     1.71 &  $<$8.17 &              39.6 &           $<$40.3 & E-CDF-S 01 \\
\\
   CXOHDFN J123631.66+620907.3 &                34.5$^{+8.5}_{-7.4}$ &               40.9$^{+10.9}_{-9.7}$ &                  1.3 &     1.09 &     2.84 &              39.8 &              40.2 &      CDF-N \\
   CXOHDFN J123632.55+621039.5 &                              $<$9.5 &                 4.4$^{+2.6}_{-2.6}$ &        1.8$^{\rm a}$ &  $<$0.32 &     0.25 &           $<$39.2 &              39.1 &      CDF-N \\
   CXOHDFN J123637.18+621135.0 &                17.8$^{+6.0}_{-4.8}$ &                21.1$^{+7.2}_{-6.0}$ &        1.8$^{\rm a}$ &     0.61 &     1.21 &              39.0 &              39.3 &      CDF-N \\
   CXOHDFN J123641.81+621132.1 &                33.6$^{+7.5}_{-6.4}$ &                39.9$^{+8.8}_{-7.6}$ &                  1.7 &     1.16 &     2.39 &              39.4 &              39.7 &      CDF-N \\
   CXOHDFN J123701.47+621845.9 &                44.2$^{+9.3}_{-8.2}$ &              50.2$^{+11.7}_{-10.6}$ &                  1.6 &     1.48 &     3.21 &              40.4 &              40.7 &      CDF-N \\
   CXOHDFN J123701.99+621122.1 &                19.3$^{+6.2}_{-5.1}$ &                21.9$^{+7.4}_{-6.2}$ &        1.8$^{\rm a}$ &     0.60 &     1.15 &              39.5 &              39.7 &      CDF-N \\
\\
   CXOHDFN J123706.12+621711.9 &                17.6$^{+6.8}_{-5.7}$ &                34.0$^{+9.6}_{-8.3}$ &        1.8$^{\rm a}$ &     0.55 &     1.80 &              40.0 &              40.5 &      CDF-N \\
   CXOHDFN J123715.94+621158.3 &                17.7$^{+6.3}_{-5.2}$ &                20.3$^{+7.8}_{-6.6}$ &        1.8$^{\rm a}$ &     0.62 &     1.21 &              39.2 &              39.5 &      CDF-N \\
   CXOHDFN J123721.60+621246.8 &                18.6$^{+6.5}_{-5.4}$ &                29.8$^{+8.7}_{-7.6}$ &        1.8$^{\rm a}$ &     0.77 &     2.07 &              39.3 &              39.8 &      CDF-N \\
   CXOHDFN J123723.45+621047.9 &                21.0$^{+7.3}_{-6.1}$ &                             $<$27.7 &        1.8$^{\rm a}$ &     0.72 &  $<$1.60 &              39.4 &           $<$39.7 &      CDF-N \\
   CXOHDFN J123727.71+621034.3 &                32.5$^{+8.6}_{-7.5}$ &              35.9$^{+11.2}_{-10.0}$ &                  1.1 &     1.07 &     2.94 &              40.1 &              40.6 &      CDF-N \\
   CXOHDFN J123730.60+620943.1 &                34.5$^{+9.1}_{-7.9}$ &              67.2$^{+13.2}_{-12.1}$ &                  0.9 &     1.15 &     6.37 &              40.5 &              41.2 &      CDF-N \\
\\
Median Values$^{\rm b}$ & 19.3 & 35.9 & 1.8$^{\rm c}$ & 1.09 & 2.84 & 39.9 & 40.2 & \ldots \\
\enddata
\tablenotetext{a}{Indicates this $\Gamma$ value was assigned; see $\S$~2.2 for details.}
\tablenotetext{b}{Quoted median values were computed using only the sources which were detected in a given band (i.e., limits were not included).}
\tablenotetext{c}{Indicates mean value determined from stacking analyses (see $\S$~2.2 for details).}
\tablecomments{All fluxes and luminosities have been corrected for Galactic absorption as discussed at the end of $\S$~1. Col.(1) Off-nuclear \hbox{X-ray} source name. Col.(2) \hbox{0.5--2.0~keV} source counts. Col.(3) \hbox{0.5--8.0~keV} source counts. Col.(4) Effective photon index $\Gamma$.  When the number of counts is low (i.e., $<$30 counts in the 0.5--2.0~keV band), we set $\Gamma = 1.8$ as per the discussion in $\S$~2.2.  Other values of $\Gamma$ are as reported by Alexander et al. (2003; CDF-N and CDF-S) or Lehmer et al. (2005; E-CDF-S).  Col.(5) \hbox{0.5--2.0~keV} flux in units of $10^{-16}$ \flux. Col.(6) \hbox{0.5--8.0~keV} flux in units of $10^{-16}$ \flux. Col.(7) Logarithm of the \hbox{0.5--2.0~keV} luminosity in units of \xlum. Col.(8) Logarithm of the \hbox{0.5--8.0~keV} luminosity in units of \xlum. Col.(9) Survey field in which each source was identified.  For E-CDF-S identifications, the associated field number (i.e., 01--04) indicates the \chandra\ pointing within which the source was detected (see Lehmer et~al. 2005 for details).}
\end{deluxetable}

\end{rotate}

%
%

\begin{rotate}

\begin{deluxetable}{lccccccccc}
\tablenum{2}
\tabletypesize{\scriptsize}
\tablewidth{0pt}
\tablecaption{Off-Nuclear Sources: Additional Properties}

\tablehead{
\colhead{}                          &
\multicolumn{3}{c}{Positional Offset In Units Of\ldots}         &
\colhead{$V_{606}$}                          &
\colhead{}                          &
\colhead{$M_{606}$}                          &
\colhead{$\nu L_{\nu}$ (6000 \AA)}                          &
\colhead{H04 Detection?}            &
\colhead{Optical Knot?}            \\

\colhead{Source Name}               &
\colhead{(arcsec)}                  &
\colhead{(pos. err.)}                &
\colhead{(kpc)}                     &
\colhead{(mag)}                     &
\colhead{$z$}                       &
\colhead{(mag)}                     &
\colhead{($\log$ cgs)}              &
\colhead{(Y/N)}                     &
\colhead{(Y/N)}                     \\

\colhead{(1)}         &
\colhead{(2)}         &
\colhead{(3)}         &
\colhead{(4)}         &
\colhead{(5)}         &
\colhead{(6)}         &
\colhead{(7)}         &
\colhead{(8)}         &
\colhead{(9)}         &
\colhead{(10)}         
}

\startdata
CXOECDFS J033122.00$-$273620.1 &  3.60 &  2.11 &  8.88 & 18.80 &    0.140$^{\rm p}$ & $-$20.1 &    43.4 & N & N \\
CXOECDFS J033128.84$-$275904.8 &  1.70 &  2.00 &  6.97 & 19.22 &    0.266$^{\rm p}$ & $-$21.4 &    43.9 & N & N \\
CXOECDFS J033139.05$-$280221.1 &  2.28 &  2.25 &  8.96 & 18.24 &    0.251$^{\rm p}$ & $-$22.2 &    44.2 & N & N \\
CXOECDFS J033143.46$-$275527.8 &  1.71 &  2.02 &  3.49 & 17.94 &    0.112$^{\rm p}$ & $-$20.4 &    43.5 & N & N \\
CXOECDFS J033143.48$-$275103.0 &  2.28 &  1.91 &  9.28 & 19.57 &    0.265$^{\rm p}$ & $-$21.1 &    43.8 & N & N \\
 CXOCDFS J033219.10$-$274445.6 &  1.21 &  2.02 &  4.01 & 20.84 &    0.201$^{\rm p}$ & $-$19.0 &    43.0 & N & N \\
\\
 CXOCDFS J033221.91$-$275427.2 &  1.81 &  2.37 &  5.58 & 19.82 &    0.184$^{\rm p}$ & $-$19.9 &    43.3 & N & N \\
 CXOCDFS J033230.01$-$274404.0 &  0.91 &  1.52 &  1.31 & 17.53 &    0.076$^{\rm s}$ & $-$20.2 &    43.4 & Y & Y \\
 CXOCDFS J033234.73$-$275533.8 &  4.71 &  5.24 &  3.55 & 16.78 &    0.038$^{\rm s}$ & $-$19.4 &    43.1 & Y & N \\
CXOECDFS J033249.26$-$273610.6 &  3.30 &  5.03 &  6.72 & 17.90 &    0.112$^{\rm p}$ & $-$18.9 &    42.9 & N & Y \\
CXOECDFS J033316.29$-$275040.7 &  3.24 &  2.31 &  5.47 & 17.84 &    0.090$^{\rm p}$ & $-$20.4 &    43.5 & N & N \\
CXOECDFS J033322.97$-$273430.7 &  2.30 &  1.51 &  4.08 & 19.20 &    0.096$^{\rm p}$ & $-$18.9 &    42.9 & N & N \\
\\
   CXOHDFN J123631.66+620907.3 &  1.26 &  2.04 &  3.29 & 20.15 &    0.150$^{\rm p}$ & $-$19.1 &    43.0 & N & Y \\
   CXOHDFN J123632.55+621039.5 &  2.02 &  3.36 &  4.86 & 20.82 &    0.136$^{\rm s}$ & $-$18.2 &    42.7 & N & Y \\
   CXOHDFN J123637.18+621135.0 &  2.23 &  3.71 &  3.31 & 18.39 &    0.079$^{\rm s}$ & $-$19.3 &    43.1 & Y & Y \\
   CXOHDFN J123641.81+621132.1 &  1.23 &  2.04 &  2.04 & 20.08 &    0.089$^{\rm s}$ & $-$17.9 &    42.6 & Y & Y \\
   CXOHDFN J123701.47+621845.9 &  2.89 &  4.65 & 10.69 & 18.98 &    0.232$^{\rm s}$ & $-$21.3 &    43.9 & Y & N \\
   CXOHDFN J123701.99+621122.1 &  1.24 &  2.07 &  2.98 & 19.45 &    0.136$^{\rm s}$ & $-$19.6 &    43.2 & Y & Y \\
\\
   CXOHDFN J123706.12+621711.9 &  1.01 &  1.68 &  3.98 & 19.07 &    0.254$^{\rm s}$ & $-$21.4 &    44.0 & N & Y \\
   CXOHDFN J123715.94+621158.3 &  1.11 &  1.84 &  2.13 & 18.70 &    0.105$^{\rm s}$ & $-$19.7 &    43.3 & Y & N \\
   CXOHDFN J123721.60+621246.8 &  2.62 &  4.36 &  5.07 & 18.15 &    0.106$^{\rm s}$ & $-$20.3 &    43.5 & Y & Y \\
   CXOHDFN J123723.45+621047.9 &  2.87 &  4.31 &  5.90 & 19.56 &    0.113$^{\rm s}$ & $-$19.0 &    43.0 & Y & N \\
   CXOHDFN J123727.71+621034.3 &  2.05 &  2.73 &  7.14 & 18.51 &    0.214$^{\rm s}$ & $-$21.5 &    44.0 & N & N \\
   CXOHDFN J123730.60+620943.1 &  1.48 &  1.68 &  6.54 & 19.48 &    0.298$^{\rm s}$ & $-$21.5 &    44.0 & N & Y \\
\\
Median Values & 2.05 & 2.11 & 5.07 & 19.07 & 0.14 & $-$20.2 & 43.4 & \ldots & \ldots \\
\enddata
\tablecomments{Col.(1) Off-nuclear \hbox{X-ray} source name. Col.(2) \hbox{X-ray} positional offset from the associated optical-source nucleus in units of arcseconds. Col.(3) Same offset in previous column in units of the \chandra\ positional error. Col.(4) Physical offset in units of kpc. Col.(5) ACS $V_{606}$ observed magnitude of host galaxy. Col.(6) Redshift estimate of the host galaxy for each off-nuclear \hbox{X-ray} source. Superscripts ``s'' and ``p'' indicate spectroscopic and photometric redshifts, respectively (see $\S$~2 for details).  Col.(7) Absolute magnitude of the host galaxy ($M_{606}$) at $\lambda \approx 6000$ \AA.  Col.(8) Logarithm of the host-galaxy, rest-frame 6000 \AA\ luminosity in units of \xlum.  Col.(9) Indicates whether each source was previously detected by H04. Col.(10) Indicates whether an optical knot is observed to be coincident with the \hbox{X-ray} source position.}
\end{deluxetable}

\end{rotate}

%
%

\begin{figure}
\figurenum{1}
\centerline{
\includegraphics[width=16.0cm]{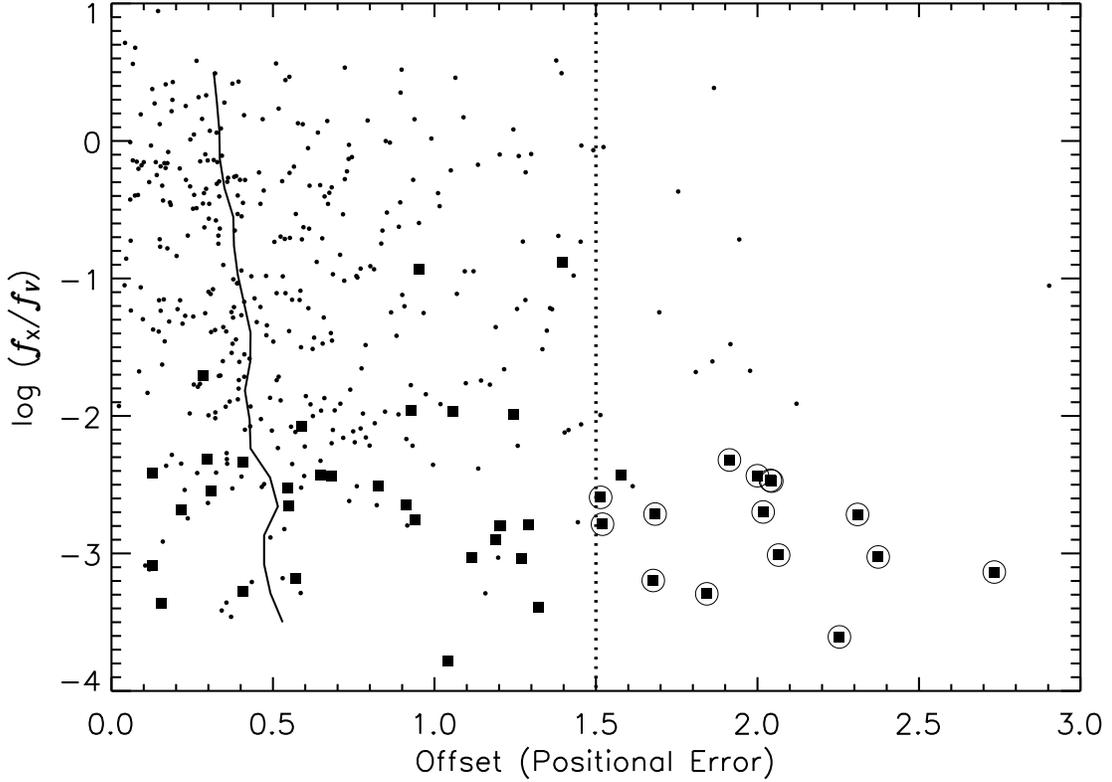}
}
\caption{\small
Logarithm of the \hbox{X-ray-to-optical} flux ratio vs. offset in units of
positional error for \hbox{0.5--2.0~keV} detected sources matched to optical
sources with $V_{606} < 24$.  Large filled squares indicate sources with
$V_{606} < 21$, and the circled sources are our off-nuclear source candidates.
Flux ratios were computed using point-source \hbox{X-ray} fluxes and integrated
host-galaxy optical fluxes.  The vertical dotted line indicates the location of
1.5 $\times$ the \chandra\ positional error, which was used to distinguish
sources as off-nuclear (see $\S$~2.1).  Sources with larger $\log [f_{\rm
X}/f_{\rm V}]$ that lie outside 1.5 $\times$ the \chandra\ positional error are
generally low-significance matches and lie outside the optical extent of the
matched source.  The solid curve indicates the running median offset for
sources within a region of size $\Delta \log [f_{\rm X}/f_{\rm V}] = 1.5$.  The
general trend of improved median positional accuracy with increasing $\log
[f_{\rm X}/f_{\rm V}]$ is expected as the number of ``point-like'' AGNs
contributing to the statistics continually increases.  
} \end{figure}

%
%

\begin{figure}
\figurenum{2}
\centerline{
\includegraphics[width=16.0cm]{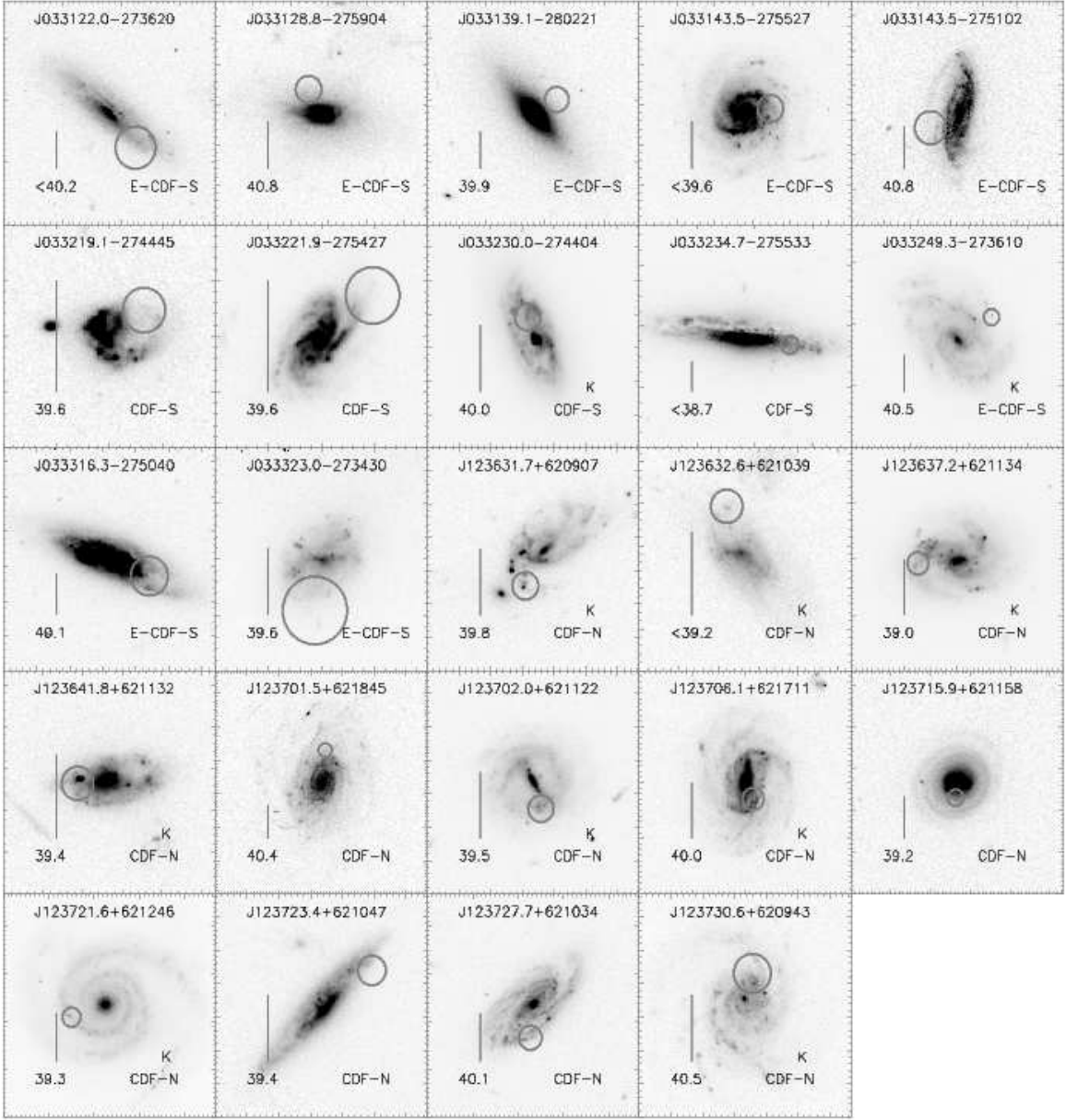}
}
\caption{\small
Advanced Camera for Surveys (ACS) $V_{606}$-band postage-stamp images of each
off-nuclear source host galaxy.  In each image, we show the off-nuclear source
name ({\it top\/}), the survey field in which the source is detected ({\it lower right\/}), and
the logarithm of the \hbox{0.5--2.0~keV} luminosity in \xlum\ ({\it lower left\/}); a
``K'' is displayed above the survey field if the off-nuclear source is
visually coincident with an optical knot.  All sources are detected in either the
\hbox{0.5--2.0~keV} or \hbox{0.5--8.0~keV} bands; the upper limits shown here
are for sources not detected in the \hbox{0.5--2.0~keV} band.  For illustrative
purposes, the images are not all the same size; the scale of each image can be
deduced from the vertical 3\arcsec\ bar located in the lower-left corner of
each frame.  Each off-nuclear source is marked with a circle having a radius
equal to the \chandra\ 80--90\% positional error.
}
\end{figure}

%
%

\begin{figure}
\figurenum{3}
\centerline{
\includegraphics[width=16.0cm]{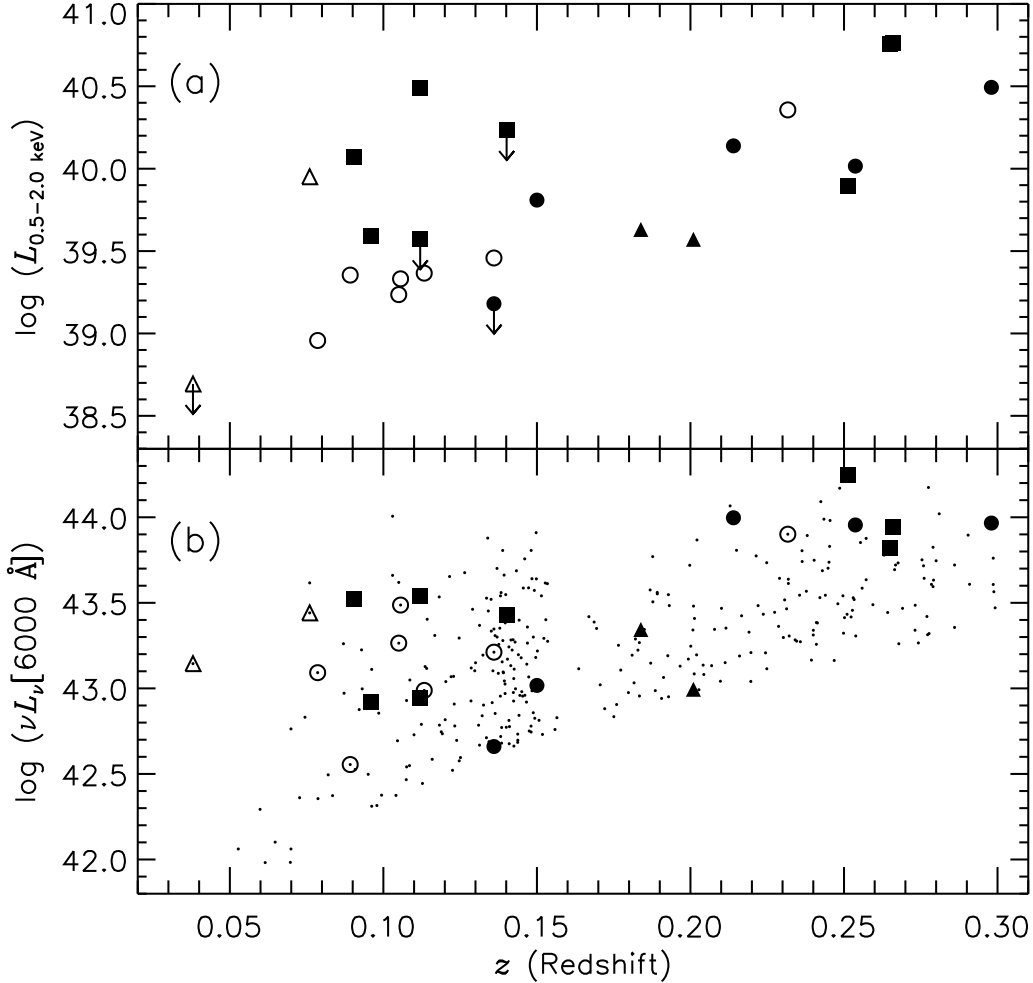}
}
\vspace{-0.5in}
\caption{
{\bf (a)} \hbox{X-ray} luminosity and redshift of each off-nuclear source.
Symbols correspond to objects detected in the \hbox{CDF-N} ({\it circles\/}),
\hbox{CDF-S} ({\it triangles\/}), and \hbox{E-CDF-S} ({\it squares\/}) surveys.
Filled symbols correspond to off-nuclear sources unique to this study (i.e.,
not previously discovered by Hornschemeier et~al. 2004).  Upper limits
correspond to sources detected in the \hbox{0.5--8.0~keV} band but not in the
\hbox{0.5--2.0~keV} band.  {\bf (b)} Rest-frame 6000 \AA\ optical luminosity
for spiral and irregular galaxies with $V_{606} < 21$ as a function of redshift
({\it small filled circles\/}).  Galaxies hosting an off-nuclear source have been
outlined with symbols following the convention of Figure~3a.} \end{figure}

%
%

\begin{figure}
\figurenum{4}
\centerline{
\includegraphics[width=16.0cm]{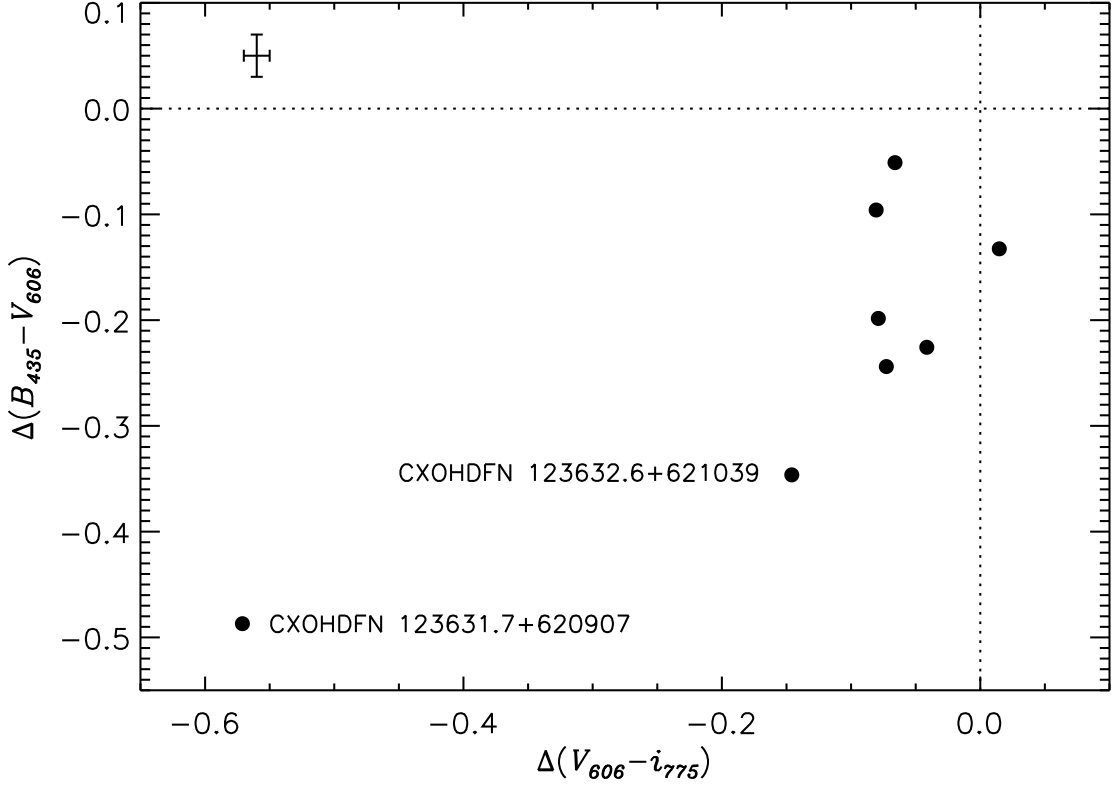}
}
\caption{
Relative color difference between optical knot and host galaxy ($\Delta
(B_{435}-V_{606})$ vs. $\Delta (V_{606}-i_{775})$; see $\S$~2.2) for the eight 
optical knots coincident with off-nuclear \hbox{X-ray} sources in the GOODS
regions ({\it filled circles\/}).  The plotted error bar in the upper-left corner shows
the typical errors of these measurements; the dotted lines show the expected
values for the case of the optical knots having the same colors as their host
galaxies.  The relatively blue optical-knot colors show that these off-nuclear
regions are likely populated by a younger stellar population; the source names
for the two most extreme cases have been noted. } \end{figure}

%
%

\begin{figure}
\figurenum{5}
\centerline{
\includegraphics[width=16.0cm]{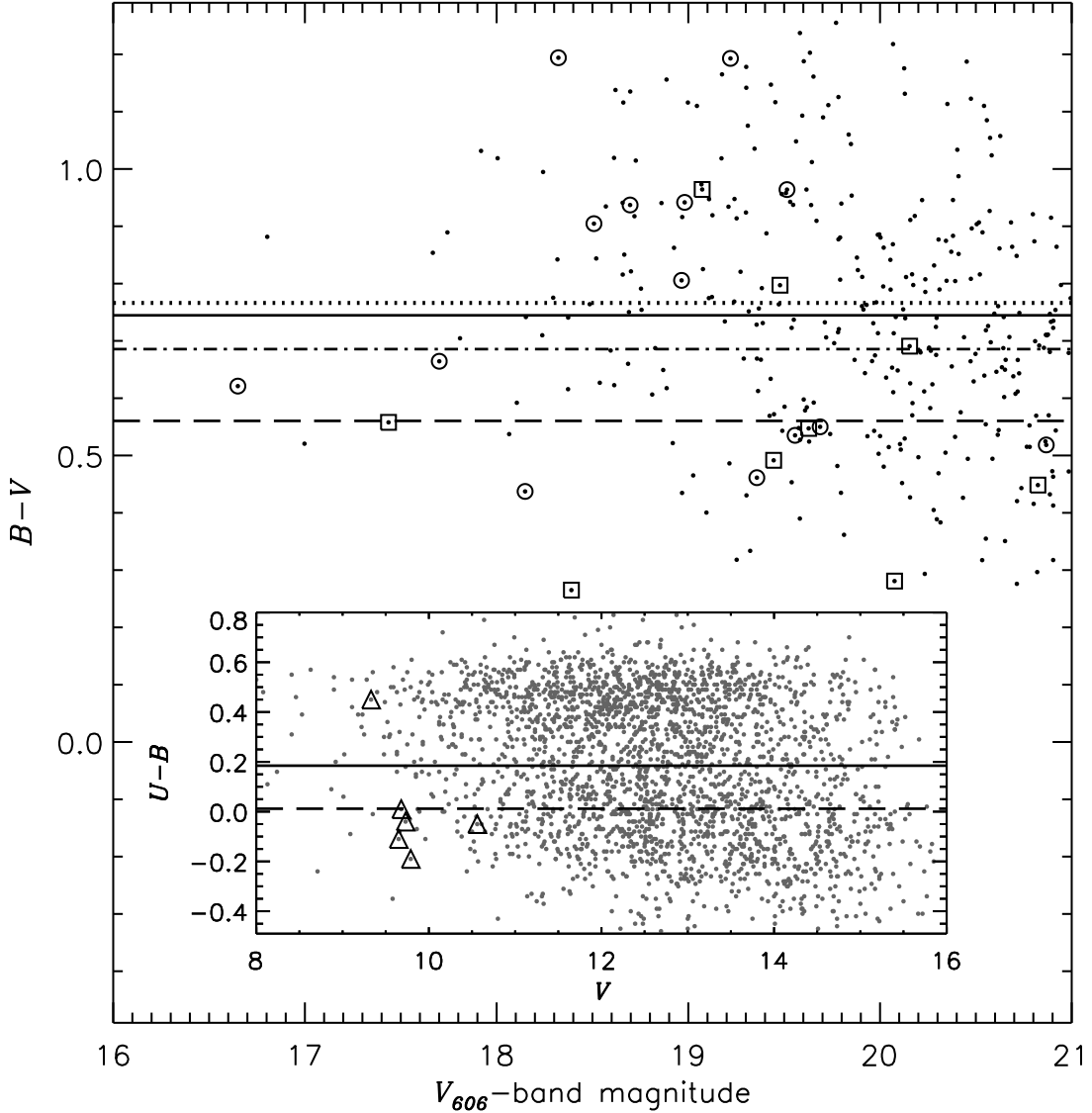}
}
\caption{
Optical $B-V$ color vs. $V_{606}$ magnitude for $V_{606} < 21$, $z < 0.3$
field galaxies in the CDFs ({\it small filled circles\/}).  Open symbols indicate
galaxies hosting off-nuclear \hbox{X-ray} sources with ({\it squares\/}) and without
({\it circles\/}) optical knots.  The horizontal lines indicate mean $B-V$ values for
field galaxies ({\it solid line\/}), galaxies hosting off-nuclear sources ({\it dot-dashed
line\/}), and galaxies hosting off-nuclear sources with ({\it dashed line\/}) and without
({\it dotted line\/}) optical knots.  The inset plot shows the optical colors (i.e.,
Johnson $U-B$ vs. $V$) for local galaxies from the RC3 catalog ({\it gray filled
circles\/}).  The six sources highlighted by open triangles are galaxies hosting
IXOs coincident with luminous optical knots, and the horizontal lines represent
the mean $U-B$ values for RC3 galaxies ({\it solid line\/}) and the subset of RC3
IXO-hosting galaxies with optical knots ({\it dashed line}\/).  Note that for
both samples, the optical knots are preferentially located in relatively blue,
star-forming galaxies.  }
\end{figure}

%
%

\begin{figure}
\figurenum{6}
\centerline{
\includegraphics[width=16.0cm]{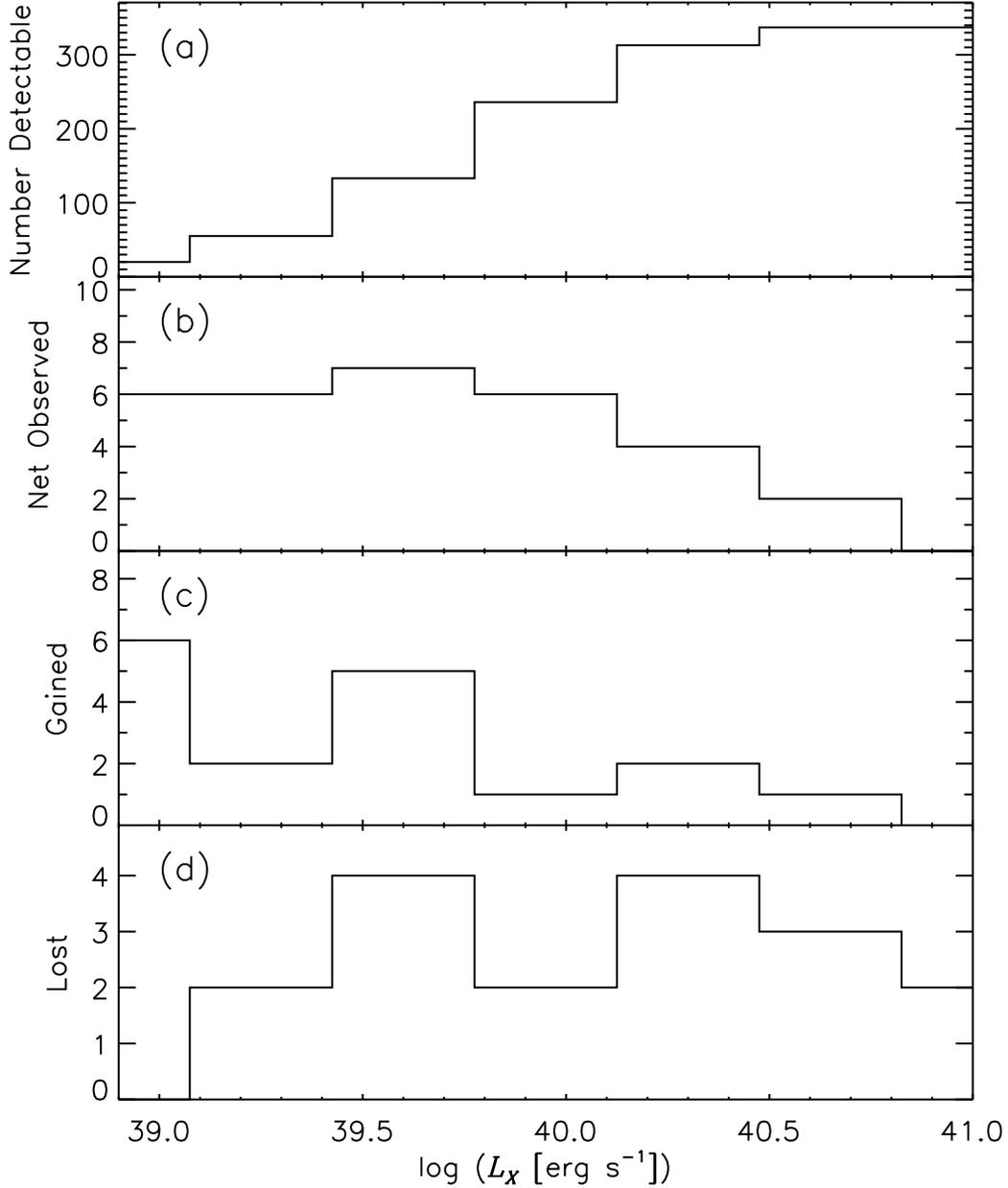}
}
\vspace{-0.5in}
\caption{
{\bf (a)} Number of galaxies in the \chandra\ deep fields in which we could
detect an off-nuclear source of \hbox{0.5--2.0~keV} luminosity \Lx.  The
highest \Lx\ bin contains all 337 spiral and irregular galaxies in our sample. {\bf
(b)} Number of galaxies in each \Lx\ bin of panel (a) containing an off-nuclear
source with an \hbox{X-ray} luminosity of \Lx\ or greater.  {\bf (c)} Number
of galaxies gained in each \Lx\ bin of panel (b) progressing from the lowest
\Lx\ bin to the highest.  {\bf (d)} Number of galaxies dropping out of each
\Lx\ bin of panel (b) progressing from the lowest \Lx\ bin to the highest.} 
\end{figure}

%
%

\begin{figure}
\figurenum{7}
\centerline{
\includegraphics[width=14.0cm]{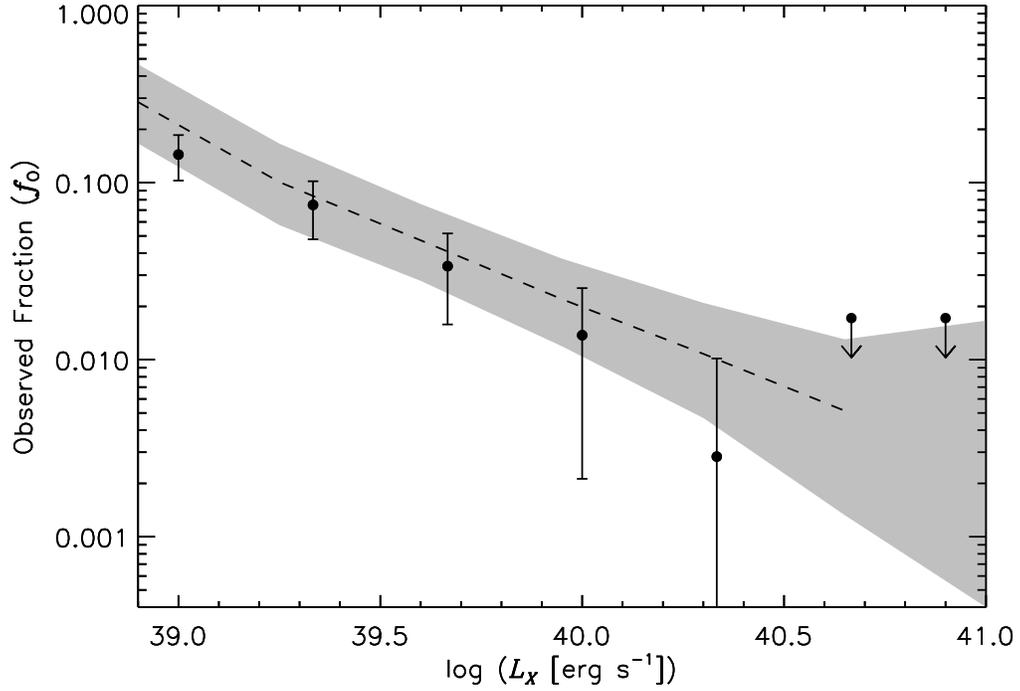}
}
\caption{
Observed fraction of galaxies in the \chandra\ deep fields 
hosting an off-nuclear source with a \hbox{0.5--2.0~keV} luminosity of \Lx\ or
greater ({\it dashed line\/}).  The shaded area shows the 1$\sigma$ confidence region,
computed using the methods outlined in Gehrels (1986).  The dashed line
terminates for \Lx~$\simgt 10^{40.7}$~\xlum\ due to the lack of off-nuclear
sources with luminosities in this regime; the shaded region here represents the
3$\sigma$ upper limit.  The solid points with 1$\sigma$ error bars and
3$\sigma$ upper limits represent the observed fraction for the matched PC04
subsample of local galaxies. } 
\end{figure}

%
%

\begin{figure}
\figurenum{8}
\centerline{
\includegraphics[width=14.0cm]{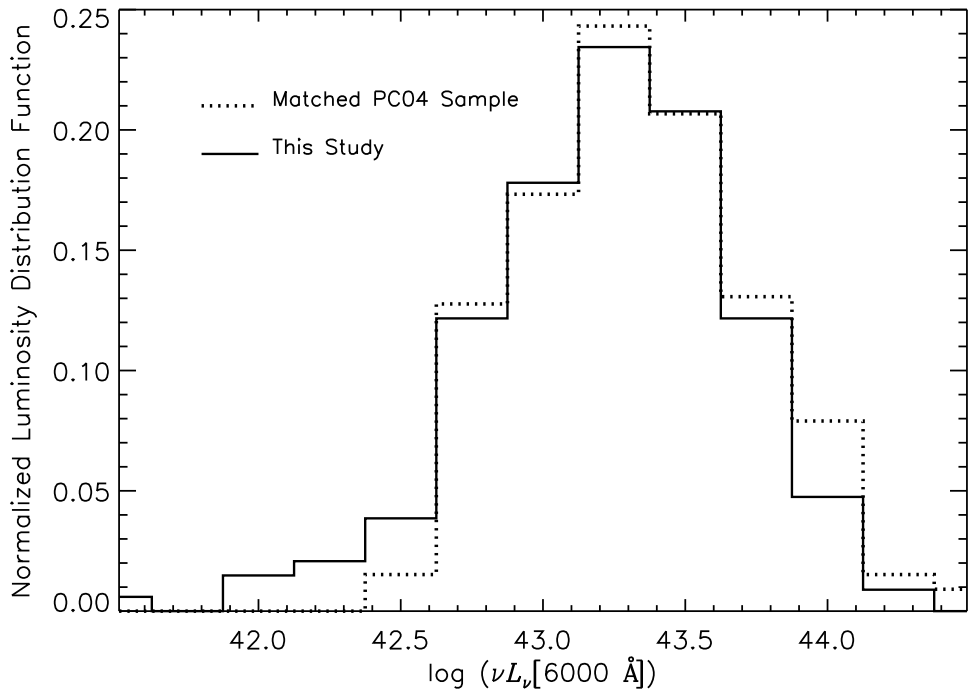}
}
\caption{
Normalized optical-luminosity distributions for field galaxies in this
survey ({\it solid histogram\/}) and a matched sample of galaxies in the local universe
adapted from the PC04 sample ({\it dotted histogram\/}).  To construct the matched
sample, we adjusted the original PC04 galaxy sample to include only galaxies with
luminosities \hbox{$\nu L_{\nu}$(6000 \AA) $\simgt 10^{42.6}$ \xlum}; this is
equivalent to requiring $V_{606} < 21$ for these galaxies if they were placed
at $z = 0.14$. }
\end{figure}

%
%

\begin{figure}
\figurenum{9}
\centerline{
\includegraphics[width=16.0cm]{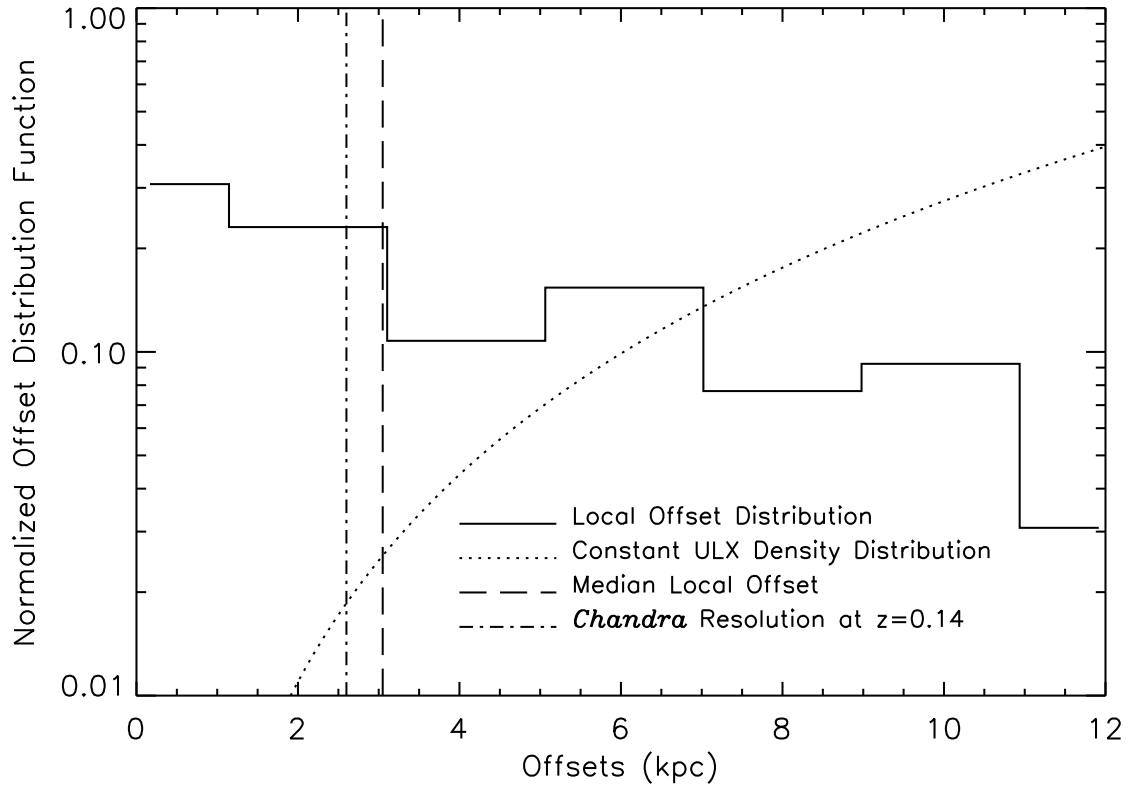}
}
\caption{Projected physical offset distribution function of \chandra-observed
local IXOs ({\it solid histogram\/}).  The dashed vertical line represents the median
offset for the IXOs used to create the offset distribution function.  The dot-dashed
vertical line indicates the approximate resolution of \chandra\ at $z = 0.14$,
the median redshift of our intermediate-redshift, off-nuclear \hbox{X-ray}
source sample.  The dotted curve shows the expected distribution of offsets for
a constant-density galactic distribution of off-nuclear \hbox{X-ray} sources.
}
\end{figure}

%
%

\begin{figure}
\figurenum{10}
\centerline{
\includegraphics[width=20.0cm]{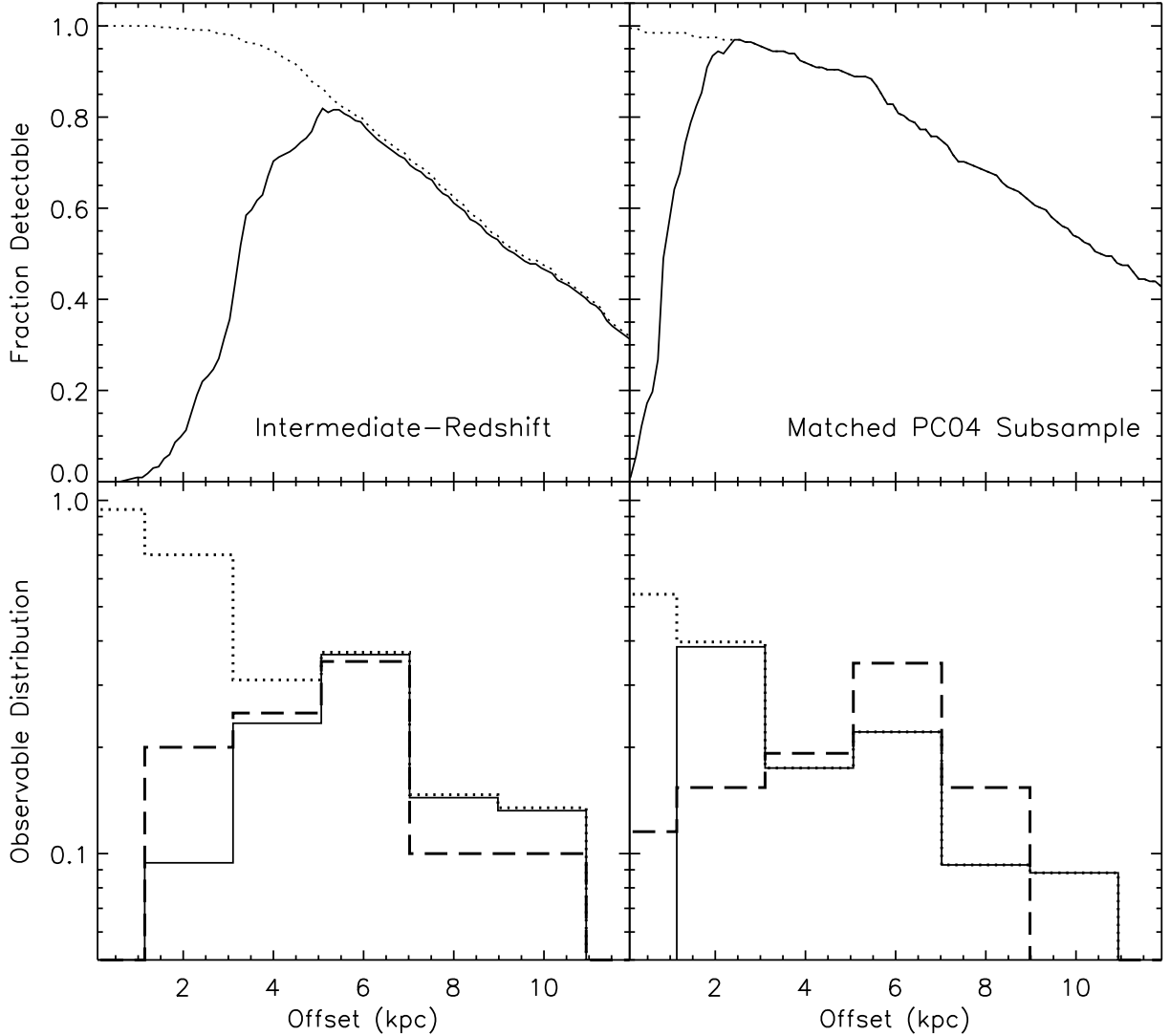}
}
\vspace{-0.5in}
\caption{{\bf (top panels)} Fraction of galaxies in which an off-nuclear
source could have been detected at a given offset for our data ({\it top-left panel\/})
and for IXOs in the PC04 sample assuming a galaxy size of 0.5 $\times$ the RC3
major-axis diameter, $D_{25}$ ({\it top-right panel\/}).  Solid lines correspond to the
actual resolution of the observations, and dotted lines correspond to the
case where there is no resolution limit. {\bf (bottom panels)} The solid
and dotted histograms show the expected distributions of off-nuclear sources
given the respective resolution constraints in the top panels and the local
\chandra-observed IXO distribution (i.e., Figure~9); the left and right panels
correspond to our data and the PC04 data, respectively.  For comparison, the
actual measured distributions have been plotted ({\it dashed histograms\/}).  
}
\end{figure}

%
%

\begin{figure}
\figurenum{11}
\centerline{
\includegraphics[width=15.0cm]{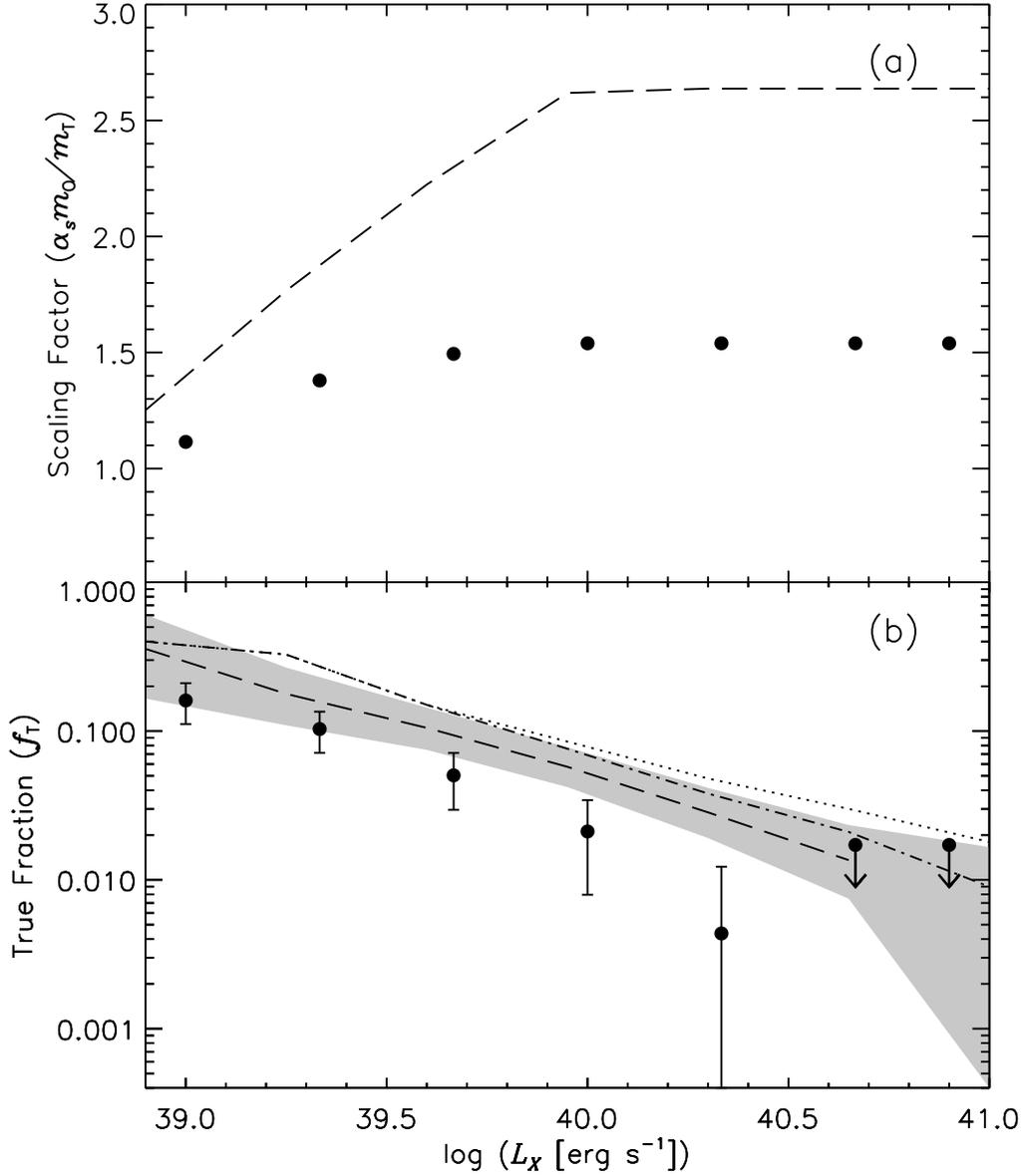}
}
\vspace{-0.5in}
\caption{\small
{\bf (a)} Scaling factor ($\alpha_{\rm s}m_{\rm O}/m_{\rm T}$) applied to the
observed fraction ($f_{\rm O}$; Figure~7) to obtain the true fraction ($f_{\rm
T}$; Figure~11b) for our sample ({\it dashed line\/}) and the matched PC04
subsample ({\it filled circles\/}).  These factors were obtained using the
methods discussed in $\S$~3.  {\bf (b)} True fraction ($f_{\rm T}$) of galaxies
in the \chandra\ deep fields hosting an off-nuclear source with a
\hbox{0.5--2.0~keV} luminosity of \Lx\ or greater ({\it dashed line with shaded
1$\sigma$ confidence region\/}).  The filled circles with error bars represent
the equivalent true fraction for the matched PC04 subsample of local galaxies.
In the region where \Lx~$\simgt 10^{40.7}$~\xlum, there are no off-nuclear
sources in either sample (i.e., the intermediate-redshift or PC04 subsample),
and therefore we have plotted 3$\sigma$ upper limits.  The dotted line shows
the total \hbox{X-ray} detection fraction for the 337 spiral and irregular
galaxies in our sample, which provides an upper limit to the true fraction of
galaxies hosting off-nuclear sources.  Similarly, the dot-dashed line shows the
total \hbox{X-ray} detection fraction for galaxies in our sample but with a
luminosity upper-bound of $10^{41.5}$~\xlum.  } \end{figure}

%
%

\begin{figure}
\figurenum{12}
\centerline{
\includegraphics[width=15.0cm]{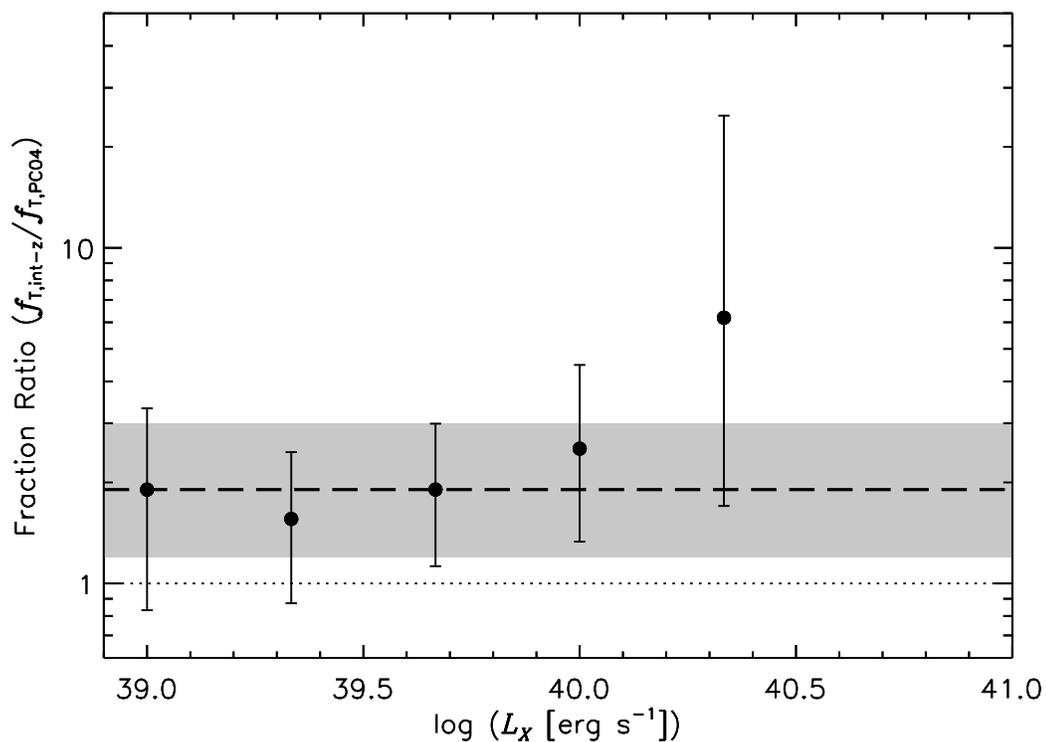}
}
\caption{
Ratios of off-nuclear source incidence fractions of our intermediate-redshift
galaxy sample and the matched local PC04 subsample ($f_{\rm T,int-{\it
z}}$/$f_{\rm T,PC04}$; {\it filled circles with 1$\sigma$ error bars\/}); the median
ratio is shown as a horizontal dashed line.  The shaded region shows the
expected increase in the fraction at \hbox{$z \approx 0.05-0.3$} due to the global
increase in star-formation density at these redshifts and the dotted horizontal
line shows the expected ratio for the case of no evolution.  }
\end{figure}


\begin{thebibliography}{} 
%

\bibitem[]{800}{} 
Alexander, D.~M., et al.\ 2003, \aj, 126, 539

\bibitem[Barger et al.(2003)]{2003AJ....126..632B} Barger, A.~J., et al.\ 
2003, \aj, 126, 632

\bibitem[]{806}{}
Bauer, F.E., Alexander, D.M., Brandt, W.N., Schneider, D.P., Treister, E.,
Hornschemeier, A.E., \& Garmire, G.P. 2004, AJ, 128, 2048

\bibitem[]{810}{}
Brandt, W.N., \& Hasinger, G. \ 2005, \araa, 43, 827

\bibitem[Capak et al.(2004)]{2004AJ....127..180C} Capak, P., et al.\ 2004, 
\aj, 127, 180

\bibitem[Colbert \& Ptak(2002)]{2002ApJS..143...25C} Colbert, E.~J.~M., \& 
Ptak, A.~F.\ 2002, \apjs, 143, 25 

\bibitem[Colbert et al.(2004)]{2004ApJ...602..231C} Colbert, E.~J.~M., 
Heckman, T.~M., Ptak, A.~F., Strickland, D.~K., \& Weaver, K.~A.\ 2004, 
\apj, 602, 231 

\bibitem[]{819}{} 
Colbert, E.~J.~M., \& Miller, M.~C. \ 2004 (astro-ph/0402677)

\bibitem[Coleman et al.(1980)]{1980ApJS...43..393C} Coleman, G.~D., Wu, 
C.-C., \& Weedman, D.~W.\ 1980, \apjs, 43, 393

\bibitem[de Vaucouleurs et al.(1991)]{1991trcb.book.....D} de Vaucouleurs, 
G., de Vaucouleurs, A., Corwin, H.~G., Buta, R.~J., Paturel, G., \& Fouque, 
P.\ 1991, Volume 1-3, XII, 2069 pp.~7 figs..~ Springer-Verlag Berlin 
Heidelberg New York

\bibitem[]{830}{}
Freeman, P.E., Kashyap, V., Rosner, R., \& Lamb, D.Q. 2002, ApJS, 138, 185

\bibitem[Ferguson et al.(2004)]{2004ApJ...600L.107F} Ferguson, H.~C., et 
al.\ 2004, \apjl, 600, L107

\bibitem[]{833}{}
Gehrels, N. 1986, ApJ, 303, 336

\bibitem[]{894}{}
Ghosh, P., White, N.~E. 2001, \apj, 559, 97

\bibitem[]{836}{}
Giacconi, R., et~al. 2002, ApJS, 139, 369 

\bibitem[]{839}{}
Giavalisco, M., et al.\ 2004, \apjl, 600, L93

\bibitem[Gilfanov et al.(2004)]{2004NuPhS.132..369G} Gilfanov, M., Grimm, 
H.-J., \& Sunyaev, R.\ 2004, Nuclear Physics B Proceedings Supplements, 
132, 369 

\bibitem[Hornschemeier et al.(2002)]{2002ApJ...568...82H} Hornschemeier, 
A.~E., Brandt, W.~N., Alexander, D.~M., Bauer, F.~E., Garmire, G.~P., 
Schneider, D.~P., Bautz, M.~W., \& Chartas, G.\ 2002, \apj, 568, 82 

\bibitem[Hornschemeier et al.(2004)]{2004ApJ...600L.147H} Hornschemeier, 
A.~E., et al.\ 2004, \apjl, 600, L147 (H04)

\bibitem[Irwin et al.(2004)]{2004ApJ...601L.143I} Irwin, J.~A., Bregman, 
J.~N., \& Athey, A.~E.\ 2004, \apjl, 601, L143

\bibitem[Kennicutt(1984)]{1984ApJ...287..116K} Kennicutt, R.~C.\ 1984, 
\apj, 287, 116

\bibitem[Kennicutt(1998)]{1998ARA&A..36..189K} Kennicutt, R.~C.\ 1998, 
\araa, 36, 189 

\bibitem[Kilgard et al.(2002)]{2002ApJ...573..138K} Kilgard, R.~E., Kaaret, 
P., Krauss, M.~I., Prestwich, A.~H., Raley, M.~T., \& Zezas, A.\ 2002, 
\apj, 573, 138 

\bibitem[King et al.(2001)]{2001ApJ...552L.109K} King, A.~R., Davies, 
M.~B., Ward, M.~J., Fabbiano, G., \& Elvis, M.\ 2001, \apjl, 552, L109

\bibitem[Lehmer et al.(2005a)]{2005ApJS..161...21L} Lehmer, B.~D., et al.\ 
2005a, \apjs, 161, 21

\bibitem[Lehmer et al.(2005b)]{2005AJ....129....1L} Lehmer, B.~D., et al.\ 
2005b, \aj, 129, 1 

\bibitem[Liu \& Bregman(2005)]{2005ApJS..157...59L} Liu, J.-F., \& Bregman, 
J.~N.\ 2005, \apjs, 157, 59

\bibitem[Liu \& Mirabel(2005)]{2005A&A...429.1125L} Liu, Q.~Z., \& Mirabel, 
I.~F.\ 2005, \aap, 429, 1125

\bibitem[Lockman(2003)]{Lockman2003}
{Lockman}, F. \ 2003, {\it Soft X-ray Emission from Clusters of Galaxies and Related Phonomena}, ed. R.~Lieu (astro-ph/0311386)

\bibitem[Loewenstein et al.(2005)]{2005ChJAA...5S.159L} Loewenstein, M., 
Angelini, L., \& Mushotzky, R.~F.\ 2005, Chinese Journal of Astronony and 
Astrophysics, 5, 159

\bibitem[]{954}{}
Lyons, L. 1991, Data Analysis for Physical Science Students.
Cambridge University Press, Cambridge

\bibitem[]{877}{}
Madau, P., Pozzetti, L., \& Dickinson, M.\ 1998, \apj, 498, 106 

\bibitem[Miller et al.(2004)]{2004ApJ...614L.117M} Miller, J.~M., Fabian, 
A.~C., \& Miller, M.~C.\ 2004, \apjl, 614, L117

\bibitem[Mobasher et al.(2004)]{2004ApJ...600L.167M} Mobasher, B., et al.\ 
2004, \apjl, 600, L167

\bibitem[Nandra et al.(2005)]{2005MNRAS.356..568N} Nandra, K., et al.\ 
2005, \mnras, 356, 568 

\bibitem[]{889}{}
Pakull, M.~W., \& Mirioni, L. \ 2002, in New Visions of the X-Ray Universe in
the \xmm\ and \chandra\ Era, ed. F. Jansen et al. (astro-ph/0202488)

\bibitem[P{\'e}rez-Gonz{\'a}lez et al.(2005)]{2005ApJ...630...82P} 
P{\'e}rez-Gonz{\'a}lez, P.~G., et al.\ 2005, \apj, 630, 82

\bibitem[Ptak \& Colbert(2004)]{2004ApJ...606..291P} Ptak, A., \& Colbert, 
E.\ 2004, \apj, 606, 291 (PC04)

\bibitem[]{958}{}
Ramsey, C.~J., Williams, R.~M., Gruendl, R.~A., Chen, C.-H.~R., Chu, Y.-H., \&
Wang, Q.~D. \ 2006, \apj, in-press, (astro-ph/0511540)

\bibitem[]{899}{}
Rix, H., et al.\ 2004, \apjs, 152, 163

\bibitem[Schiminovich et al.(2005)]{2005ApJ...619L..47S} Schiminovich, D., 
et al.\ 2005, \apjl, 619, L47

\bibitem[Spergel et al.(2003)]{Spergel2003} 
Spergel, D.~N., et al.\ 2003, \apjs, 148, 175

\bibitem[Stark et al.(1992)]{Stark1992} Stark, A.~A., Gammie,
C.~F., Wilson, R.~W., Bally, J., Linke, R.~A., Heiles, C., \& Hurwitz, M.\
1992, \apjs, 79, 77

\bibitem[Swartz et al.(2004)]{2004ApJS..154..519S} Swartz, D.~A., Ghosh, 
K.~K., Tennant, A.~F., \& Wu, K.\ 2004, \apjs, 154, 519

\bibitem[Szokoly et al.(2004)]{2004ApJS..155..271S} Szokoly, G.~P., et al.\ 
2004, \apjs, 155, 271 

\bibitem[Wirth et al.(2004)]{2004AJ....127.3121W} Wirth, G.~D., et al.\ 
2004, \aj, 127, 3121

\bibitem[]{921}{}
Wolf, C., et al.\ 2004, \aap, 421, 913


%
\end{thebibliography}
\end{document}